# Design Principles for High Temperature Superconductors with Hydrogen-based Alloy Backbone at Moderate Pressure

Zihan Zhang[1], Tian Cui[2, 1, *], Michael J. Hutcheon[3], Alice M. Shipley[3], Hao Song[1], Mingyang Du[1], Vladimir Z. Kresin[4], Defang Duan[1, †], Chris J. Pickard[5, 6], Yansun Yao[7]

[1]State Key Laboratory of Superhard Materials, College of Physics, Jilin University, Changchun 130012, China
[2]Institute of High Pressure Physics, School of Physical Science and Technology, Ningbo University, Ningbo, 315211, China
[3]Theory of Condensed Matter Group, Cavendish Laboratory, University of Cambridge, J. J. Thomson Avenue, Cambridge CB3 0HE, United Kingdom
[4]Lawrence Berkeley Laboratory, University of California at Berkeley, Berkeley, CA 94720, USA
[5]Department of Materials Science & Metallurgy, University of Cambridge, 27 Charles Babbage Road, Cambridge CB3 0FS, United Kingdom
[6]Advanced Institute for Materials Research, Tohoku University 2-1-1 Katahira, Aoba, Sendai, 980-8577, Japan
[7]Department of Physics and Engineering Physics, University of Saskatchewan, Saskatoon, Saskatchewan S7N 5E2, Canada

Corresponding authors email: [*]cuitian@nbu.edu.cn, cuitian@jlu.edu.cn
[†]duandf@jlu.edu.cn

# Abstract :

Hydrogen-based superconductors provide a route to the long-sought goal of room-temperature superconductivity, but the high pressures required to metallize these materials limit their immediate application. For example, carbonaceous sulfur hydride, the first room-temperature superconductor, can reach a critical temperature ($T_c$) of 288 K only at the extreme pressure of 267 GPa. The next recognized challenge is the realization of room-temperature superconductivity at significantly lower pressures. Here, we propose a strategy for the rational design of high-temperature superconductors at low pressures by alloying small-radius elements and hydrogen to form ternary hydride superconductors with alloy backbones. We identify a hitherto unknown 'fluorite-type' backbone in compositions of the form $AXH_8$, which exhibit high temperature superconductivity at moderate pressures. The $Fm\bar{3}m$ phase of $LaBeH_8$, with a 'fluorite-type' H-Be alloy backbone, is predicted to be metastable and superconducting with a $T_c \sim$ 191 K at 50 GPa; a substantially lower pressure than that required by the geometrically similar clathrate hydride $LaH_{10}$ (170 GPa). Our approach paves the way for finding high-$T_c$ ternary hydride superconductors at conditions close to ambient pressures.

Hydrogen, the lightest element, has been predicted to become a metallic solid and exhibit high-temperature superconductivity (with $T_c$s in the range 100-760 K) under extreme pressures [1]. However, metallization of solid hydrogen is still uncertain in high-pressure experiments up to 490 GPa [2]. It was predicted that comparable high-temperature superconductivity could be achieved in hydrogen dominant materials by "chemically pre-compressing" the hydrogen with other elements to produce the valence density sufficient for metallization at lower pressures [3]. Based on this principle, a series of H-based superconductors were theoretically predicted and some were successfully synthesized in the laboratory. Notably, $H_3S$ was predicted to be a high-temperature superconductor with a $T_c$ of 191-204 K [4], which was later confirmed by an experimentally measured $T_c$ of 203 K at 155 GPa [5,6]. Following this success, several new hydrides in the clathrate hydride family, which consist of a pure hydrogen backbone pre-compressed by heavy metal atoms, were predicted and then synthesized, including $LaH_{10}$ with a $T_c$ of 250-260 K at 170-180 GPa [7-11]. Several geometric classes of hydrides were found to facilitate high $T_c$; in addition to the covalent six-fold cubic $H_3S$ and the sodalite-type clathrate hydrides, a class of "penta-graphene-like" hydrides with high $T_c$ were recently predicted at 250 GPa [12]. Although the pressures at which these H-based superconductors become stable are much lower than pure metallic hydrogen, the required pressure (> 150 GPa) is still difficult to obtain. The next challenge is therefore the realization of room-temperature superconductivity at significantly lower pressures (with a clear final goal of reaching ambient pressure).

Various routes have been explored to reduce the pressure at which hydride superconductors become stable. Doping known hydrogen-rich binary systems with extra elements or molecules is one way to achieve this. For example, the ionic bonds formed by doping a $H_3S$ host with $CH_4$ molecules lead to stability of the compound at a much lower pressure of 100 GPa [13,14]. A careful choice of the elements used for pre-compression is also important. For example, low-pressure stability in lanthanide and actinide systems correlates strongly with the presence of $f$ states at the Fermi level. As a result, metastable phases of $YbH_6$ and $LuH_6$ are predicted to exhibit high-$T_c$ superconductivity at relatively low pressures (145 K at 70 GPa and 273 K at 100 GPa, respectively [15]). Turning to even lower pressures, $Fm\overline{3}m$ $UH_{8+\delta}$ [16], $F\overline{4}3m$ $EuH_9$ [17] and $C2/c$ $NdH_7$ [18] have been observed experimentally at 42 GPa, 86 GPa and 85 GPa, respectively, but with very low critical temperatures.

Whilst binary hydrides have been extensively explored [19-23], research into ternary hydrides is much more challenging. Because phase diagrams of ternary A-X-H systems that is much more complex than those of binary systems. Despite this difficulty, many ternary hydrides have been found to exhibit favorable properties when compared to current binary systems. For example, C doped $H_3S$ possesses a much higher $T_c$ than that of H$_3$S in experiments [24], and Li$_2$MgH$_{16}$, a molecular Mg-H phase doped with Li, is predicted to have the highest $T_c$ to date (473 K) [25]. The ability to methodically and efficiently explore ternary hydrides and identify those with desirable superconducting properties at low pressures is key to advancing research in superconductivity.

Here, we propose a strategy to design high-$T_c$ ternary hydride superconductors at low pressures by engineering binary X-H backbones, which are subsequently "pre-compressed" by a large-radius element A. The resulting alloyed backbones are easier to metallize compared to pure H backbones, such as those found in the clathrate hydrides. These backbones can be designed by doping familiar structures with additional atoms (X), which break the local motif of the parent structure. This leads to a metallic H-rich phase with occupied overlapping bands, known as a *hydrogen alloy* phase [3]. The IIA and IIIB metals, with large radii, are known to be effective "pre-compressors" [7,8,26], so we investigate their combination with these designer backbones. We use a hard-sphere model to investigate the stability of these new materials in terms of the radius and bond lengths, leading to the prediction of a new class of high-$T_c$ hydride superconductors with a 'fluorite-type' backbone.

We design the first of these alloy backbone materials from the pure H backbone of the high-temperature superconductor LaH$_{10}$ [10], which possesses the same symmetry as the low-pressure UH$_8$ superconductor [16]. Comparing the crystal structures of UH$_8$ and LaH$_{10}$ (as shown in Fig. 1(a)-(b)), the structure of LaH$_{10}$ can be regarded as a UH$_8$ parent structure doped with additional H atoms at vacant tetrahedral sites (as shown in Fig. 1(d)). The extra H atoms break the localization of cubic H$_8$ units and lead to the famous clathrate backbone in LaH$_{10}$. The extra dopant X atoms with small-radius can instead be located in vacancy sites at the centre of the cubic H$_8$ units, and the result is an $Fm\bar{3}m$ structure of LaXH$_8$. For example, atom Be may be suitable dopants at these cubic sites. This novel H-Be alloy backbone corresponds to a fluorite-type arrangement, in which Be atoms are located on the sites of a face-centered cubic lattice, and [H$_4$] tetrahedra are present in the tetrahedral vacancies found between the Be atoms (see Figs. 1(e)-(f)). This results in a novel class of ternary hydrides AXH$_8$, designed with X-H alloy backbones to potentially achieve high-temperature superconductivity at low pressure. We note that during the preparation of our manuscript, this same cubic structure was reported in the La-B-H ternary system [27,28].

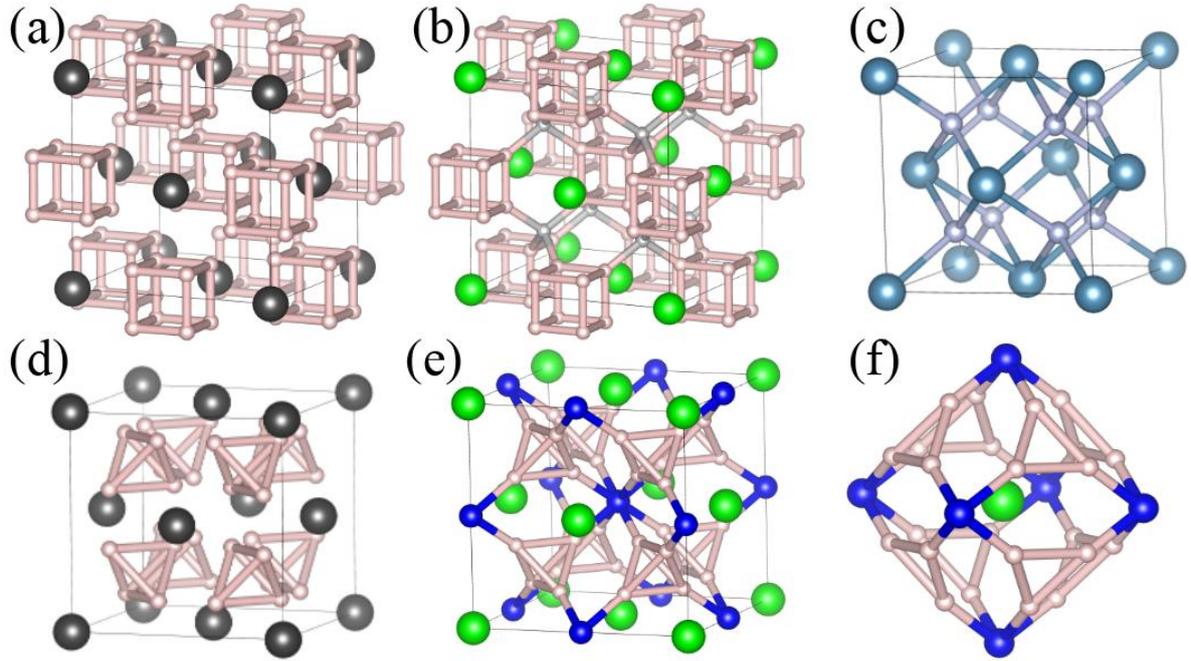

FIG. 1 The parent structure and two types of doped framework. (a) The crystal structure of $UH_8$, with [$H_8$] cubic units shown. The U ions are shown in black and the H ions in pink. (b) The crystal structure of $LaH_{10}$, in which La ions are shown in green. The backbone in $LaH_{10}$ consists of cubic-unit H atoms (pink) and tetrahedron-center H atoms (gray). (c) The crystal structure of fluorite $CaF_2$, the Ca cations are shown in dark blue and the F ions in light blue. (d) The crystal structure of $UH_8$, with [$H_4$] tetrahedral units shown. (e) The crystal structure of $LaBeH_8$, in which La ions are shown in green. The backbone in $LaBeH_8$ consists of tetrahedral-unit H atoms (pink) and cubic-center Be atoms (blue). (f) The 'fluorite-type' cage of $LaBeH_8$, consisting of Be-H bonds and H-H bonds.

Having identified this fluorite-type ternary structure, we go on to investigate a wider class of fluorite-type backbone hydrides $AXH_8$ ($Fm\bar{3}m$) formed from "pre-compressor" elements A (A = Sc, Ca, Y, Sr, La, Ba) and small-radius elements X (X = Be, B, Al). The phonon dispersions of these $AXH_8$ structures were calculated in the pressure range 50-200 GPa, among which 7 compounds, $CaBH_8$, $CaBeH_8$, $YBeH_8$, $SrBH_8$, $LaAlH_8$, $LaBH_8$ and $LaBeH_8$, were found to be dynamically stable (as shown in Fig 2 and Fig. S26-S32). $LaBeH_8$ is dynamically stable at 50 GPa, a moderate pressure accessible to Kawai-Type Multi-Anvil presses (KMAPs) [29]. Four other compounds were found to be dynamically stable at or below 100 GPa; $LaBH_8$ (70 GPa), $LaAlH_8$ (100 GPa), $CaBH_8$ (100 GPa) and $YBeH_8$ (100 GPa). The other two compounds, $SrBH_8$ and $CaBeH_8$, become dynamically stable at 150 GPa and 210 GPa, respectively. We go on to determine the thermodynamic stability of these dynamically stable "fluorite-like" backbone hydrides using *ab initio* random structure searching (AIRSS) [30], by constructing convex hulls (see Figs. S5-S11). Four compounds, LaBeH$_8$, CaBH$_8$, YBeH$_8$ and SrBH$_8$, sit on the convex hull. LaBeH$_8$ is thermodynamically stable at 98 GPa. CaBH$_8$, YBeH$_8$ and SrBH$_8$ become thermodynamically stable at, or above, 300 GPa as shown in the SM [31].

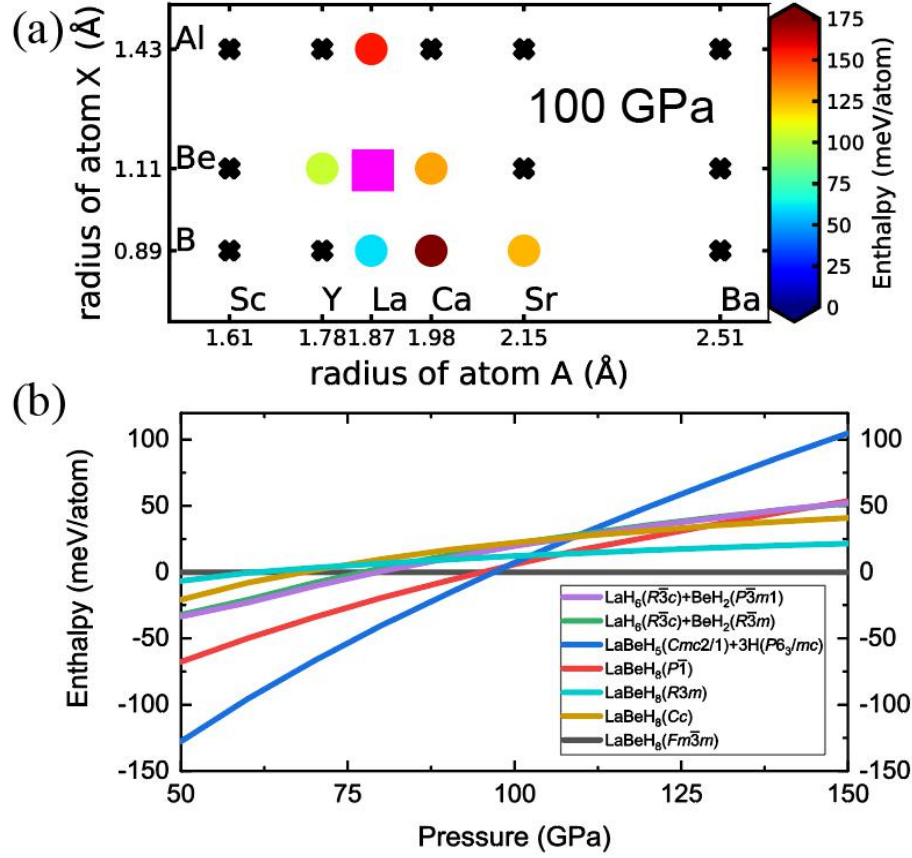

FIG. 2 Calculated enthalpy of alloyed ternary hydrides AXH$_8$ above the convex hull. (a) The radius of atom A is plotted on the x-axis and the radius of atom X on the y-axis. Dynamically unstable systems are shown as black crosses. Metastable phases are shown as circles, colored according to the calculated enthalpy above the convex hull. Thermodynamically stable phases are shown as carmine squares. (b) Calculated enthalpy as a function of pressure for La-Be-H structures relative to the $Fm\bar{3}m$ phase of LaBeH$_8$, where structures of LaH$_6$, BeH$_2$ and H are from Refs. [7-9,48,53], respectively.

Having investigated the stability of these fluorite-type backbone hydrides, we go on to investigate their superconducting properties. Based on Eliashberg equations, the values of $T_c$ were determined using $\mu^* = 0.1$. As shown in Fig. 3, LaBeH$_8$ demonstrates high-temperature superconductivity with a $T_c$ of 191 K at a remarkably low pressure of 50 GPa. This is possible because the local structure of the H$_8$ molecular units is broken by the Be dopants. These fluorite-type backbone structures can even exhibit room-temperature superconductivity, with the metastable hydride CaBeH$_8$ predicted to possess a $T_c$ of 300 K at 210 GPa. Likewise, YBeH$_8$, CaBH$_8$, LaBH$_8$,

LaAlH$_8$ and SrBH$_8$ also exhibit high-temperature superconductivity with 249 K at 100 GPa, 238 K at 100 GPa, 160 K at 70 GPa, 144 K at 100 GPa and 200 K at 150 GPa, respectively. As can be seen in Fig. 3, the threshold pressure at which fluorite-type backbone hydrides become dynamically stable is lower than that for typical high-$T_c$ hydrides, whilst retaining a $T_c$ that is much higher than the temperature of liquid nitrogen. LaBeH$_8$ is the first proposed H-based superconductor with a figure of merit [54] score around $S = 3$.

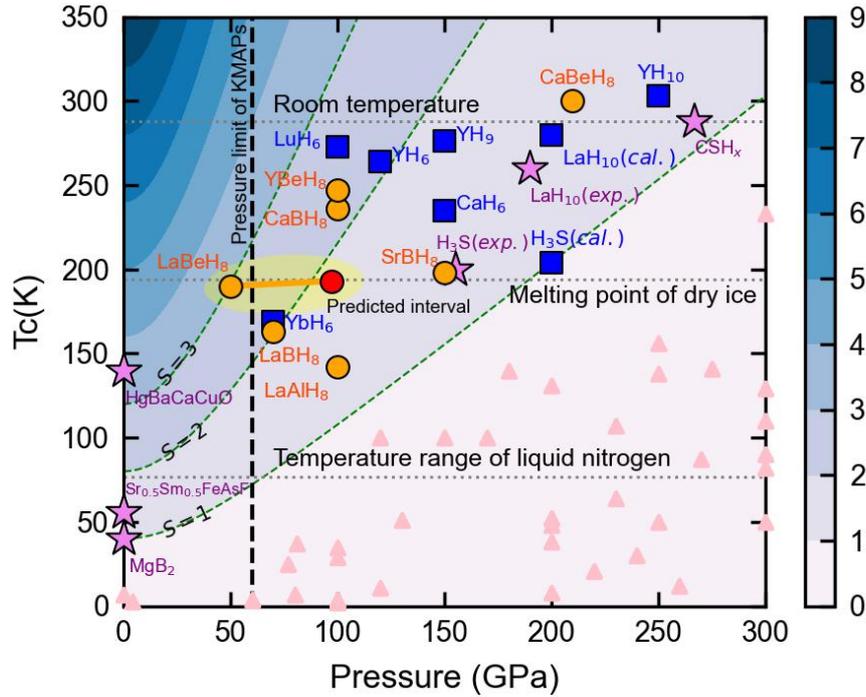

FIG. 3 Pressure dependence of $T_c$s calculated for alloyed clathrate hydrides shown alongside other high-Tc superconductors. The orange circles are $T_c$s of fluorite-type backbone hydrides at the lowest pressure where they become dynamically stable. The red circle is at the lowest pressure where LaBeH$_8$ becomes thermodynamically stable (98 GPa), and the suggested synthesis pressure range for cubic LaBeH$_8$ is highlighted in yellow. The blue squares are $T_c$s of clathrate binary hydrides at the lowest pressures reported in Refs. [4,7,8,26,54]. The purple stars are $T_c$s of well-known superconductors from experiment [5,10,24,54]. The background is shaded according to the figure of merit $S = \dfrac{T}{\sqrt{P^2 + T_{MgB_2}^2}}$ used to evaluate the significance of a particular superconductor [54].

The dynamic stability of fluorite-type hydrides depends on the radii of the pre-compressor element A as shown in Fig. 2(a), suggesting that a hard spheres model [55] derived from the geometry may allow us to draw general conclusions about this novel structure family. To construct this model, we make two simplifications: i) the large-radius atoms A are regarded as hard spheres; ii) the backbone is characterized by H-H bonds and X-H bonds. The derivation of this model is presented in the SM [31] and the solution gives the lattice parameter of the cubic unit cell ($L$), the lengths of H-H bonds ($b_{H-H}$) and X-H bonds ($b_{X-H}$) as follows:

$$\begin{cases} L = \frac{\sqrt{3}+1}{t+1} \cdot 2R_A \\ b_{H-H} = \frac{\sqrt{3}+1-2t}{t+1} \cdot \sqrt{6}R_A = A_{H-H}d_{H-H}, \\ b_{X-H} = \frac{3t-\sqrt{3}}{t+1} \cdot R_A = A_{H-X}d_{H-X} \end{cases}$$

where $R_A$ is the covalent radius [56] of atom A and t = 0.95-1.05 is a tolerance factor allowing slight overlap of the hard spheres. The bond lengths can also be represented as products of flexible factors ($A$) and bond lengths of binary hydrides ($d$) from the literatures.

Table. 1 The lattice parameter $L$, bond lengths $b_{X-H}$ and $b_{H-H}$ and flexible factors $A$ calculated from the hard-spheres model (unprimed) and from DFT (primed) at 150 GPa. The amount of charge transferred to H is denoted as $\delta$. Here we use $t$=1.03, $d_{H-H}$ =1.1 Å [7,12,25], $d_{Be-H}$ = 1.31 Å [48], $d_{B-H}$ = 1.22 Å [50] and $d_{Al-H}$ = 1.72 Å [51,57].

| | $L$, Å | $b_{X-H}$, Å ($A_{X-H}$) | $b_{H-H}$, Å ($A_{H-H}$) | $L'$, Å | $b'_{X-H}$, Å ($A'_{X-H}$) | $b'_{H-H}$, Å ($A'_{H-H}$) | $\delta$, e-/atom |
|---|---|---|---|---|---|---|---|
| LaBeH$_8$ | 5.03 | 1.25 (91%) | 1.52 (137%) | 5.17 | 1.36 (99%) | 1.43 (130%) | 0.36 |
| LaBH$_8$ | 5.03 | 1.25 (102%) | 1.52 (137%) | 5.13 | 1.33 (109%) | 1.45 (131%) | 0.32 |
| LaAlH$_8$ | 5.03 | 1.25 (73%) | 1.52 (137%) | 5.38 | 1.52 (88%) | 1.32 (120%) | 0.46 |
| SrBH$_8$ | 5.79 | 1.44 (118%) | 1.74 (158%) | 5.05 | 1.32 (108%) | 1.41 (128%) | 0.29 |
| CaBH$_8$ | 5.33 | 1.32 (109%) | 1.60 (146%) | 4.89 | 1.3 (107%) | 1.33 (121%) | 0.29 |
| CaBeH$_8$ | 5.33 | 1.32 (97%) | 1.60 (146%) | 4.93 | 1.32 (96%) | 1.32 (120%) | 0.32 |
| YBeH$_8$ | 4.79 | 1.19 (86%) | 1.44 (131%) | 5.02 | 1.34 (98%) | 1.36 (124%) | 0.40 |

The solutions for the 7 hydrides considered in this work are shown in Table 1, alongside with the values calculated from DFT at 150 GPa. The value of $L$ obtained in the hard spheres model is similar to that calculated by DFT in the pressure range of 100-200 GPa. The geometry of the fluorite-type backbone has elongated H-H bond lengths by 30-60% or 20-30% (according to the hard-spheres model and DFT, respectively) compared to the H-H bond lengths in common hydrides [7,12,25], because of the large amount of charge transferred to the H-H bond (Table 1). The X-H bonds are also affected by the geometry, but whether the bonds are elongated or shortened depends on the composition; the hard-spheres model and DFT calculations agree on the trend.

The hard-spheres model captures the general characteristics of AXH$_8$ and allows us to determine the criteria for elements that can be substituted for X in the crystal. In particular, candidate elements should form bonds to hydrogen with lengths in the range of 1.2-1.6 Å in binary systems. Therefore, we investigate the possibility of substituting Si, P and S into LaXH$_8$ since the X-H bond lengths are $d_{Si-H}$ ~1.6 Å [58], $d_{P-H}$ ~1.4 Å [59] and $d_{S-H}$ ~1.5 Å [60]. We find that LaSiH$_8$, LaSH$_8$ and LaPH$_8$ are all dynamically stable high-$T_c$ superconductors with 150 K at 100 GPa, 150 K at 200 GPa and 180 K at 200 GPa, respectively.

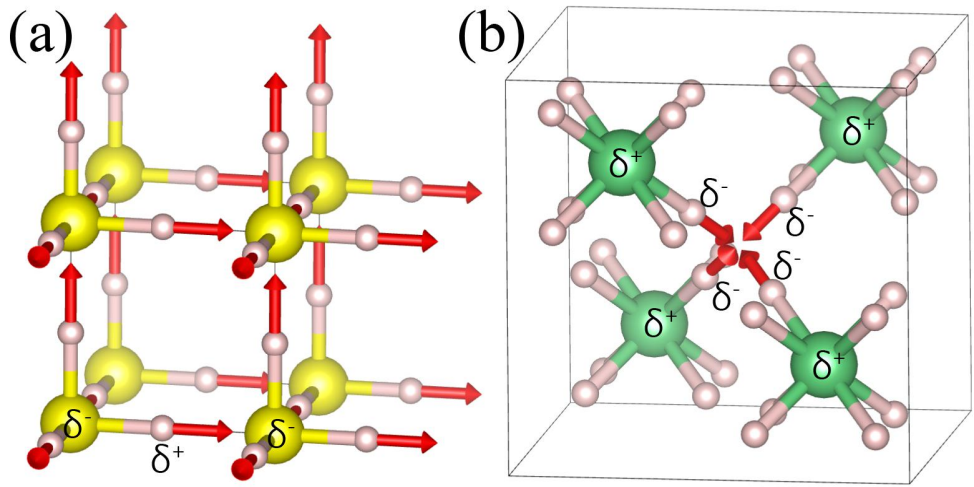

FIG. 4 The structures of [H$_3$S] and [BeH$_8$] frameworks with polarization vectors shown. (a) The polarization vectors of H-S bonds are aligned, in these bonds the positive charge is located at H atoms (pink) and negative charge is located at S atoms (yellow). (b) The polarization vectors of Be-H bonds converge, leading to a concentration of charge at the hydrogen atoms. In these bonds the positive charge is located at Be atoms (green) and the negative charge is located at H atoms (pink).

Unlike pure hydrogen frameworks, alloyed backbones in hydrides typically consist of polar bonds at high pressure. For example, the [SH$_3$] backbone of SCH$_7$ consists of H-S polar bonds as shown in Fig. 4(a). Similarly to hydrogen bonded structures at ambient pressure, end-to-end arrangements of polar bonds in these alloy backbones generally have low enthalpy. In contrast to the end-to-end arrangements found in common hydrides, in the [BeH$_8$] backbone, four H atoms with negative charge δ$^-$ are found in close proximity and form a regular tetrahedron as shown in Fig. 4(b). This distribution of charge suggests that the fluorite-type backbone is formed in a different way to the common covalent hydrides. In the unique bonding environment of the fluorite-type backbone, a large amount of charge transfer to the H atoms is possible as shown in Table 1; these electrons occupy anti-bonding states along the H-H bonds (Fig. S19-S25). The electrons in the anti-bonding orbitals not only support the elongated H-H bonds, but also give rise to an increased H-derived density of states at the Fermi level. The unique chemistry in alloyed frameworks therefore provides a route to improved H-based superconductors.

As part of efforts to develop new H-based superconductors that are stable at low pressure, we design a new class of ternary hydrides $AXH_8$ with fluorite-type alloy backbones. The design of these fluorite-type hydrides is envisioned using a hard-spheres model, which allows the creation of new fluorite-type hydrides by geometry. In this work, we investigate a total of 10 different fluorite-type hydrides to demonstrate this design, which turns out to be highly successful. All predicted hydrides show both excellent superconductivity and low-pressure dynamic stability compared with other reported H-based superconductors. Although the lowest pressure at which any of these hydrides becomes dynamically stable (~ 50 GPa) is still high when compared to ambient pressure, it is decreased by a factor of 4 from the pressure needed to stabilize typical hydride superconductors. One of the systems we considered, $LaBeH_8$, is dynamically stable down to pressures obtainable in KMAPs. Synthesis of a high-temperature hydride superconductor in a KMAP would represent a breakthrough in the field. The hydrides in the fluorite-type backbone family would number about 20 if other elements in the lanthanide and actinide series were included. Future investigations of other ternary systems may identify high-temperature superconductors even closer to ambient pressure by similar methods.

In summary, designing hydrides with an alloyed backbone is shown to be an effective approach to obtaining low-pressure H-based superconductors. In these materials, small radius elements are alloyed with hydrogen in order to introduce new bonds into the backbone instead of pure H-H bonds. These new bonds are more stable than H-H bonds, and the resulting fluorite-type backbones can be stabilized at much lower pressures. Our results provide an effective method for the rational design of lower-pressure H-based superconductors and stimulate further experimental exploration of hydride superconductors at close-to-ambient pressures. The synthesis of hydride-based superconductors at closer to ambient pressures in experiment would represent an important breakthrough in the field of high-temperature superconductivity.

## Acknowledgements

This work was supported by the National Natural Science Foundation of China (Grants No. 11974133, No. 51632002, No. 11674122), National Key R&D Program of China (Grant No. 2018YFA0305900), Program for Changjiang Scholars and Innovative Research Team in University (Grant No. IRT_15R23), and the Natural Sciences and Engineering Research Council of Canada (NSERC). C.J.P. acknowledges financial support from the Engineering and Physical Sciences Research Council (Grant EP/P022596/1) and a Royal Society Wolfson Research Merit award. AMS is funded by an EPSRC studentship. MJH acknowledges the EPSRC Centre for Doctoral Training in Computational Methods for Materials Science for funding under grant number EP/L015552/1. Some of the calculations were performed at the High Performance Computing Center of Jilin University and using TianHe-1(A) at the National Supercomputer Center in Tianjin.

[1] E. Wigner and H. B. Huntington, *On the Possibility of a Metallic Modification of Hydrogen,* J. Chem. Phys. **3**, 764, (1935).
[2] P. Dalladay-Simpson, R. T. Howie, and E. Gregoryanz, *Evidence for a new phase of dense hydrogen above 325 gigapascals,* Nature **529**, 63, (2016).
[3] N. W. Ashcroft, *Hydrogen dominant metallic alloys: High temperature superconductors?,* Phys. Rev. Lett. **92**, 187002, (2004).
[4] D. Duan, Y. Liu, F. Tian, D. Li, X. Huang, Z. Zhao, H. Yu, B. Liu, W. Tian, and T. Cui, *Pressure-induced metallization of dense $(H_2S)_2H_2$ with high-Tc superconductivity,* Sci. Rep. **4**, 6968, (2014).
[5] A. P. Drozdov, M. I. Eremets, I. A. Troyan, V. Ksenofontov, and S. I. Shylin, *Conventional superconductivity at 203 kelvin at high pressures in the sulfur hydride system,* Nature **525**, 73, (2015).
[6] M. Einaga, M. Sakata, T. Ishikawa, K. Shimizu, M. I. Eremets, A. P. Drozdov, I. A. Troyan, N. Hirao, and Y. Ohishi, *Crystal structure of the superconducting phase of sulfur hydride,* Nature Phys. **12**, 835, (2016).
[7] H. Liu, I. I. Naumov, R. Hoffmann, N. W. Ashcroft, and R. J. Hemley, *Potential high-T-c superconducting lanthanum and yttrium hydrides at high pressure,* Proc. Natl. Acad. Sci. U. S. A. **114**, 6990, (2017).
[8] F. Peng, Y. Sun, C. J. Pickard, R. J. Needs, Q. Wu, and Y. M. Ma, *Hydrogen Clathrate Structures in Rare Earth Hydrides at High Pressures: Possible Route to Room-Temperature Superconductivity,* Phys. Rev. Lett. **119**, 107001, (2017).
[9] A. M. Shipley, M. J. Hutcheon, M. S. Johnson, R. J. Needs, and C. J. Pickard, *Stability and superconductivity of lanthanum and yttrium decahydrides,* Phys. Rev. B **101**, 224511, (2020).
[10] A. P. Drozdov, P. P. Kong, V. S. Minkov, S. P. Besedin, M. A. Kuzovnikov, S. Mozaffari, L. Balicas, F. F. Balakirev, D. E. Graf, V. B. Prakapenka, E. Greenberg, D. A. Knyazev, M. Tkacz, and M. I. Eremets, *Superconductivity at 250 K in lanthanum hydride under high pressures,* Nature **569**, 528, (2019).
[11] I. Errea, F. Belli, L. Monacelli, A. Sanna, T. Koretsune, T. Tadano, R. Bianco, M. Calandra, R. Arita, F. Mauri, and J. A. Flores-Livas, *Quantum crystal structure in the 250-kelvin superconducting lanthanum hydride,* Nature **578**, 66, (2020).
[12] H. Xie, Y. Yao, X. Feng, D. Duan, H. Song, Z. Zhang, S. Jiang, S. A. T. Redfern, V. Z. Kresin, C. J. Pickard, and T. Cui, *Hydrogen Pentagraphenelike Structure Stabilized by Hafnium: A High-Temperature Conventional Superconductor,* Phys. Rev. Lett. **125**, 217001 (2020).
[13] Y. Sun, Y. Tian, B. Jiang, X. Li, H. Li, T. Iitaka, X. Zhong, and Y. Xie, *Computational discovery of a dynamically stable cubic SH3-like high-temperature superconductor at 100 GPa via CH4 intercalation,* Phys. Rev. B **101**, 174102 (2020).
[14] W. Cui, T. Bi, J. Shi, Y. Li, H. Liu, E. Zurek, and R. J. Hemley, *Route to high-T-c superconductivity via CH4-intercalated H3S hydride perovskites,* Phys. Rev. B **101**, 134504 (2020).
[15] H. Song, Z. Zhang, T. Cui, C. J. Pickard, V. Z. Kresin, and D. Duan, *High Tc superconductivity in heavy Rare Earth Hydrides: correlation between the presence of the f states on the Fermi surface, nesting and the value of Tc,* arXiv:2101.01315, (2020).
[16] I. A. Kruglov, A. G. Kvashnin, A. F. Goncharov, A. R. Oganov, S. S. Lobanov, N. Holtgrewe, S. Jiang, V. B. Prakapenka, E. Greenberg, and A. V. Yanilkin, *Uranium polyhydrides at moderate pressures: Prediction, synthesis, and expected superconductivity,* Sci. Adv. **4**, 7, (2018).
[17] D. V. Semenok, D. Zhou, A. G. Kvashnin, X. Huang, M. Galasso, I. A. Kruglov, A. G. Ivanova, A. G. Gavriliuk, W. Chen, N. V. Tkachenko, A. I. Boldyrev, I. Troyan, A. R. Oganov, and T. Cui, *Novel Strongly Correlated Europium Superhydrides,* J. Phys. Chem. Lett. **12**, 32, (2021).
[18] D. Zhou, D. V. Semenok, H. Xie, X. L. Huang, D. F. Duan, A. Aperis, P. M. Oppeneer, M. Galasso, A. I. Kartsev, A. G. Kvashnin, A. R. Oganov, and T. Cui, *High-Pressure Synthesis of Magnetic Neodymium Polyhydrides,* J. Am. Chem. Soc. **142**, 2803, (2020).
[19] J. A. Flores-Livas, L. Boeri, A. Sanna, G. Profeta, R. Arita, and M. Eremets, *A perspective on conventional high-temperature superconductors at high pressure: Methods and materials,* Phys. Rep. **856**, 1, (2020).
[20] M. J. Hutcheon, A. M. Shipley, and R. J. Needs, *Predicting novel superconducting hydrides using machine learning approaches,* Phys. Rev. B **101**, 144505 (2020).
[21] D. F. Duan, Y. X. Liu, Y. B. Ma, Z. Shao, B. B. Liu, and T. Cui, *Structure and superconductivity of hydrides at high pressures,* Natl. Sci. Rev. **4**, 121, (2017).

[22] A. R. Oganov, C. J. Pickard, Q. Zhu, and R. J. Needs, *Structure prediction drives materials discovery,* Nat. Rev. Mater. **4**, 331, (2019).

[23] E. Zurek and T. G. Bi, *High-temperature superconductivity in alkaline and rare earth polyhydrides at high pressure: A theoretical perspective,* J. Chem. Phys. **150**, 13, (2019).

[24] E. Snider, N. Dasenbrock-Gammon, R. McBride, M. Debessai, H. Vindana, K. Vencatasamy, K. V. Lawler, A. Salamat, and R. P. Dias, *Room-temperature superconductivity in a carbonaceous sulfur hydride,* Nature **586**, 373, (2020).

[25] Y. Sun, J. Lv, Y. Xie, H. Liu, and Y. Ma, *Route to a Superconducting Phase above Room Temperature in Electron-Doped Hydride Compounds under High Pressure,* Phys. Rev. Lett. **123**, 097001 (2019).

[26] H. Wang, S. T. John, K. Tanaka, T. Iitaka, and Y. Ma, *Superconductive sodalite-like clathrate calcium hydride at high pressures,* Proc. Natl. Acad. Sci. U. S. A. **109**, 6463, (2012).

[27] S. D. Cataldo, C. Heil, W. v. d. Linden, and L. Boeri, *LaBH8: the first high-Tc low-pressure superhydride,* arXiv e-prints, arXiv:2102.11227, (2021).

[28] S. D. Cataldo, W. v. d. Linden, and L. Boeri, *La-X-H hydrides: is hot superconductivity possible?,* arXiv e-prints, arXiv:2106.07266, (2021).

[29] T. Ishii, Z. D. Liu, and T. Katsura, *A Breakthrough in Pressure Generation by a Kawai-Type Multi-Anvil Apparatus with Tungsten Carbide Anvils,* Engineering **5**, 434, (2019).

[30] C. J. Pickard and Needs, *Ab initio random structure searching,* J. Phys.-Condes. Matter, 053201 (2011).

[31] *Supplemental Material,* (Includes Refs. [8,26,28,30–50].).

[32] M. D. Segall, P. J. D. Lindan, M. J. Probert, C. J. Pickard, P. J. Hasnip, S. J. Clark, and M. C. Payne, *First-principles simulation: ideas, illustrations and the CASTEP code,* J. Phys.-Condes. Matter **14**, 2717, (2002).

[33] G. Kresse and J. Furthmüller, *Efficiency of ab-initio total energy calculations for metals and semiconductors using a plane-wave basis set,* Comput. Mater. Sci. **6**, 15, (1996).

[34] J. P. Perdew and Y. Wang, *Accurate and simple analytic representation of the electron-gas correlation energy,* Phys. Rev. B **45**, 13244, (1992).

[35] J. P. Perdew, K. Burke, and Y. Wang, *Generalized gradient approximation for the exchange-correlation hole of a many-electron system,* Phys. Rev. B **54**, 16533, (1996).

[36] G. Henkelman, A. Arnaldsson, and H. Jónsson, *A fast and robust algorithm for Bader decomposition of charge density,* Comput. Mater. Sci. **36**, 354, (2006).

[37] W. Tang, E. Sanville, and G. Henkelman, *A grid-based Bader analysis algorithm without lattice bias,* J. Phys.-Condes. Matter **21**, 084204, (2009).

[38] R. Dronskowski and P. E. Bloechl, *Crystal orbital Hamilton populations (COHP): energy-resolved visualization of chemical bonding in solids based on density-functional calculations,* J. Phys. Chem. **97**, 8617, (1993).

[39] V. L. Deringer, A. L. Tchougréeff, and R. Dronskowski, *Crystal Orbital Hamilton Population (COHP) Analysis As Projected from Plane-Wave Basis Sets,* J. Phys. Chem. A **115**, 5461, (2011).

[40] P. Giannozzi, S. Baroni, N. Bonini, M. Calandra, R. Car, C. Cavazzoni, D. Ceresoli, G. L. Chiarotti, M. Cococcioni, I. Dabo, A. Dal Corso, S. De Gironcoli, S. Fabris, G. Fratesi, R. Gebauer, U. Gerstmann, C. Gougoussis, A. Kokalj, M. Lazzeri, L. Martin-Samos, N. Marzari, F. Mauri, R. Mazzarello, S. Paolini, A. Pasquarello, L. Paulatto, C. Sbraccia, S. Scandolo, G. Sclauzero, A. P. Seitsonen, A. Smogunov, P. Umari, and R. M. Wentzcovitch, *QUANTUM ESPRESSO: A modular and open-source software project for quantum simulations of materials,* J. Phys.-Condes. Matter **21**, 395502, (2009).

[41] G. Kresse and D. Joubert, *From ultrasoft pseudopotentials to the projector augmented-wave method,* Phys. Rev. B **59**, 1758, (1999).

[42] C. R. and Dynes, *McMillan's equation and the Tc of superconductors,* Solid State Commun. **10**, 615, (1972).

[43] Hertel, *Transition temperature of strong-coupled superconductors,* Phys. Rev. B **167**, 331, (1968).

[44] G. M. Eliashberg, *Interactions between electrons and lattice vibrations in a superconductor,* Sov. Phys. **11**, 696, (1960).

[45] L. P. Gor'kov and V. Z. Kresin, *Pressure and high-T-c superconductivity in sulfur hydrides,* Sci Rep **6**, 7, (2016).

[46] L. P. Gor'kov and V. Z. Kresin, *Colloquium: High pressure and road to room temperature superconductivity,* Rev. Mod. Phys. **90**, 011001 (2018).

[47] P. B. Allen and R. C. Dynes, *Transition temperature of strong-coupled superconductors reanalyzed,* Phys. Rev. B **12**, 905, (1975).

[48] Z. Wang, Y. Yao, L. Zhu, H. Liu, T. Iitaka, H. Wang, and Y. Ma, *Metallization and superconductivity of $BeH_2$ under high pressure,* J. Chem. Phys. **140**, 124707, (2014).

[49] Z. Shao, D. Duan, Y. Ma, H. Yu, H. Song, H. Xie, D. Li, F. Tian, B. Liu, and T. Cui, *Unique Phase Diagram and Superconductivity of Calcium Hydrides at High Pressures,* Inorg. Chem. **58**, 2558, (2019).

[50] C. H. Hu, A. R. Oganov, Q. Zhu, G. R. Qian, G. Frapper, A. O. Lyakhov, and H. Y. Zhou, *Pressure-induced stabilization and insulator-superconductor transition of BH,* Phys. Rev. Lett. **110**, 165504, (2013).

[51] I. Goncharenko, M. I. Eremets, M. Hanfland, J. S. Tse, M. Amboage, Y. Yao, and I. A. Trojan, *Pressure-induced hydrogen-dominant metallic state in aluminum hydride,* Phys. Rev. Lett. **100**, 045504, (2008).

[52] Y. Wang, H. Wang, J. S. Tse, T. Iitaka, and Y. Ma, *Structural morphologies of high-pressure polymorphs of strontium hydrides,* Phys. Chem. Chem. Phys. **17**, 19379, (2015).

[53] C. J. Pickard and R. J. Needs, *Structure of phase III of solid hydrogen,* Nature Phys. **3**, 473, (2007).

[54] C. J. Pickard, I. Errea, and M. I. Eremets, *Superconducting Hydrides Under Pressure,* Annu. Rev. Condens. Matter Phys. **11**, 57, (2020).

[55] A. V. Smirnov, S. G. Ponomarev, V. P. Tarasovskii, V. V. Rybal'chenko, A. A. Vasin, and V. V. Belov, *Hard-Sphere Close-Packing Models: Possible Applications for Developing Promising Ceramic and Refractory Materials,* GLASS CERAM+ **75**, 345, (2019).

[56] R. Heyrovska, *Atomic, Ionic and Bohr Radii Linked via the Golden Ratio for Elements of Groups 1 - 8 Including Lanthanides and Actinides,* Inter. J. Sci. **2**, (2013).

[57] C. J. Pickard and R. J. Needs, *Metallization of aluminum hydride at high pressures: A first-principles study,* Phys. Rev. B **76**, 144114, (2007).

[58] C. J. Pickard and R. J. Needs, *High-pressure phases of silane,* Phys. Rev. Lett. **97**, 045504 (2006).

[59] A. Shamp, T. Terpstra, T. Bi, Z. Falls, P. Avery, and E. Zurek, *Decomposition Products of Phosphine Under Pressure: PH2 Stable and Superconducting?,* J. Am. Chem. Soc. **138**, 1884, (2016).

[60] D. Duan, X. Huang, F. Tian, D. Li, H. Yu, Y. Liu, Y. Ma, B. Liu, and T. Cui, *Pressure-induced decomposition of solid hydrogen sulfide,* Phys. Rev. B **91**, 180502(R), (2015).

Supplemental Materials

for

# Design Principles for High Temperature Superconductors with Hydrogen-based Alloy Backbone at Moderate Pressure

Zihan Zhang[1], Tian Cui[2,1,*], Michael J. Hutcheon[3], Alice M. Shipley[3], Hao Song[1], Mingyang Du[1], Vladimir Z. Kresin[4], Defang Duan[1,*], Chris J. Pickard[5,6], Yansun Yao[7]

[1]*State Key Laboratory of Superhard Materials, College of Physics, Jilin University, Changchun 130012, China*
[2]*Institute of High Pressure Physics, School of Physical Science and Technology, Ningbo University, Ningbo, 315211, People's Republic of China*
[3]*Theory of Condensed Matter Group, Cavendish Laboratory, University of Cambridge, J. J. Thomson Avenue, Cambridge CB3 0HE, United Kingdom*
[4]*Lawrence Berkeley Laboratory, University of California at Berkeley, Berkeley, CA 94720, USA*
[5]*Department of Materials Science & Metallurgy, University of Cambridge, 27 Charles Babbage Road, Cambridge CB3 0FS, United Kingdom*
[6]*Advanced Institute for Materials Research, Tohoku University 2-1-1 Katahira, Aoba, Sendai, 980-8577, Japan*
[7]*Department of Physics and Engineering Physics, University of Saskatchewan, Saskatoon, Saskatchewan S7N 5E2, Canada*

# 1. The hard-sphere model of "fluorite-like" backbone hydride

We built a hard-sphere model to describe "fluorite-like" backbone hydrides. The structures of our predicted fluorite-type hydrides are similar to perovskites and alloys, so we consult their method. In the view of perovskites, ionic radius is the main factor of the stability. For alloys, three main factors are atomic radii, electronegativity and valence of elements. In case of fluorite-type hydrides $AXH_8$, atomic radii of A, the length of X-H and H-H bonds, electronegativity are main factor in our model.

Here, the model of fluorite-type hydrides $AXH_8$ were based on two assumptions: (i) the geometric models of $AXH_8$ were carried out with atomic radius of atom A and bond lengths of X-H and H-H; (ii) the electronegativity of X elements were described by transferred charge from X atoms at pressure.

According to (i) and the structure of $AXH_8$, we can carry out the three equations:

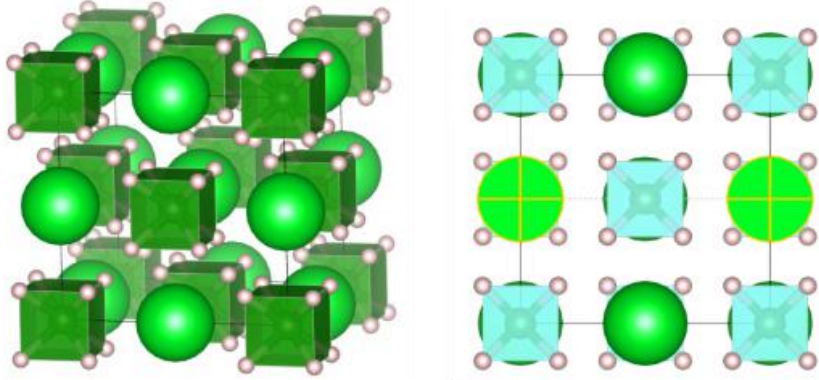

FIG. S1. (100) surface of $AXH_8$

As shown in Fig. S1, length of a side $L$ consists of two radii of hard spheres $R_A$ and a side of cubic $XH_8$ $\frac{2\sqrt{3}}{3}b_{H-X}$ (where $b_{H-X}$ is a half body diagonal of cubic $XH_8$). The equation is shown as following:

$$t \times L = 2R_A + \frac{2\sqrt{3}}{3}b_{H-X}, \qquad (S1)$$

where $L$ is the length of side of cubic unit cell, $R_A$ is the covalent radius of atom A from ref. [2], $b_{X-H}$ and $b_{H-H}$ are lengths of X-H bonds and H-H bonds. Here $t$ is tolerance factor with value around 0.95~1.05. Because the cubic $XH_8$ is not hard, it could overlap a few with hard spheres.

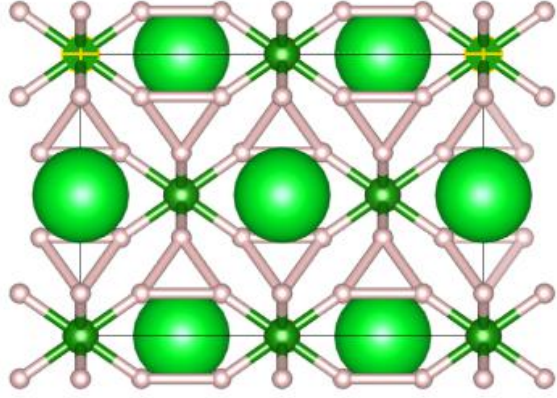

FIG. S2. (110) surface of $AXH_8$

As shown in Fig. S2, diagonal of the face (100) $\sqrt{2}L$ consists of two diagonals of cubic $XH_8$ $2\times\frac{2\sqrt{6}}{3}b_{H-X}$ and two H-H bonds $2b_{H-H}$. The equation is shown as following:

$$\frac{\sqrt{2}}{2}L = \frac{2\sqrt{6}}{3}b_{H-X} + b_{H-H}. \qquad (S2)$$

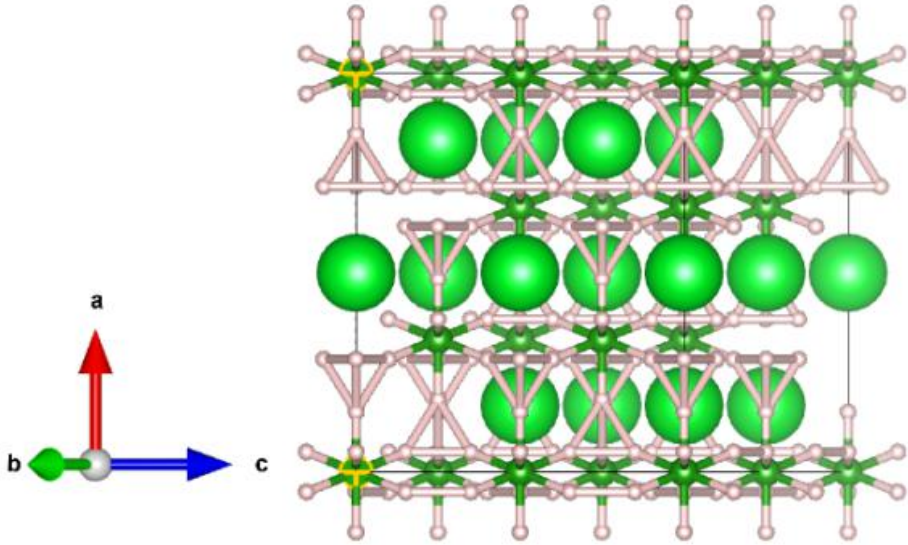

FIG. S3. ($\bar{2}11$) surface of $AXH_8$, here a is [111] and c is [$01\bar{1}$].

As shown Fig. S3, body diagonal of unit cell $\sqrt{3}L$ consists of two bonds X-H $2b_{H-X}$, two heights of H tetrahedron $\frac{2\sqrt{6}}{3}b_{H-H}$ and two radii of hard spheres $2R_A$. The equation is shown as following:

$$\frac{\sqrt{3}}{2}L = R_A + \frac{\sqrt{6}}{3}b_{H-H} + b_{H-X}, \qquad (S3)$$

Solutions of these equations are $L$, $b_{X-H}$, and $b_{H-H}$, they are side length of unit cell, lengths of H-X and H-H bonds, which are only depend on the radii of elements $R_A$ in our model as shown in table S1 (these lengths are all Å).

Table S1| The side length of unit cell $L$ ($L'$), bond lengths of H-X $b_{X-H}$ ($b'_{X-H}$) and H-H $b_{X-H}$ ($b'_{X-H}$) suggested by hard sphere model (calculated by DFT at 150 GPa). $R_A$ is the covalent radius of atom A from ref.[2]

|  | $R_A$, Å | $L$, Å | $b_{X-H}$, Å | $b_{H-H}$, Å | $L'$, Å | $b'_{X-H}$, Å | $b'_{H-H}$, Å |
|---|---|---|---|---|---|---|---|
| LaBeH$_8$ | 1.87 | 5.03 | 1.25 | 1.52 | 5.17 | 1.36 | 1.43 |
| LaBH$_8$ | 1.87 | 5.03 | 1.25 | 1.52 | 5.13 | 1.33 | 1.45 |
| LaAlH$_8$ | 1.87 | 5.03 | 1.25 | 1.52 | 5.38 | 1.52 | 1.32 |
| SrBH$_8$ | 2.15 | 5.79 | 1.44 | 1.74 | 5.05 | 1.32 | 1.41 |
| CaBH$_8$ | 1.98 | 5.33 | 1.32 | 1.60 | 4.89 | 1.3 | 1.33 |
| CaBeH$_8$ | 1.98 | 5.33 | 1.32 | 1.60 | 4.93 | 1.32 | 1.32 |
| YBeH$_8$ | 1.78 | 4.79 | 1.19 | 1.44 | 5.02 | 1.34 | 1.36 |

$L$ is the suggested length of unit cell by hard-sphere model without pressure, it is function of one variable of $R_A$. From the table, we can see that the length of unit cell, bond lengths of H-X and H-H are similar to that at pressure of 100~200 GPa by DFT calculation. It is suggested that the AXH$_8$ is a pressure-stabilized hydride.

H-H bonds of H tetrahedron in AXH$_8$ were found to be longer than those in other hydrides at pressure such as those in clathrate hydrides (~ 1.1 Å), because of more transferred charge on hydrogen atoms as shown in Table S2. In order to compare bond length with other binary hydrides, the flexible factors $A$ are used to describe the changes of lengths of H-H and X-H bonds. The bond lengths ($b$) in AXH$_8$ represented by products of flexible factors $A$ and bond lengths ($d$) in known binary hydrides and atomic phase of pure hydrogen, such that $b_{X-H} = A_{X-H} \cdot d_{X-H}$ and $b_{H-H} = A_{H-H} \cdot d_{H-H}$.

Table S2| The length of H-H bonds $b_{H-H}$ and its flexible factors $A_{H-H}$ suggested by model and calculated by DFT. The positive charge means that charge transfers to H. Here $b_{H-H} = A_{H-H} \cdot d_{H-H}$, $d_{H-H}$ is the nearest length of H-H bonds in common length of hydrides, here we use $d_{H-H}$ =1.1 Å.

|        | $b_{H-H}$, Å | $A_{H-H}$ | $b'_{H-H}$, Å | $A'_{H-H}$ | transferred charge to H, electrons |
|--------|--------------|-----------|---------------|------------|-------------------------------------|
| LaBeH$_8$ | 1.52 | 137% | 1.43 | 130% | 0.36 |
| LaBH$_8$  | 1.52 | 137% | 1.45 | 131% | 0.32 |
| LaAlH$_8$ | 1.52 | 137% | 1.32 | 120% | 0.46 |
| SrBH$_8$  | 1.74 | 158% | 1.41 | 128% | 0.29 |
| CaBH$_8$  | 1.60 | 146% | 1.33 | 121% | 0.29 |
| CaBeH$_8$ | 1.60 | 146% | 1.32 | 120% | 0.32 |
| YBeH$_8$  | 1.44 | 131% | 1.36 | 124% | 0.40 |

The lengths of H-H bonds calculated by model and by DFT are about 30~60% and 20~30% longer than the common length of hydrides. Our result shows that it is the geometry that forces the H-H bonds to take longer lengths than other, and more charge makes the long bonds stable.

To find the replaceability of element X in AXH$_8$, it is suggested to study the electronegativity of X. Here, we studied the transferred charge from X atoms instead of electronegativity of element X.

Here we used common minimal X-H bond lengths $d_{X-H}$ of BeH$_2$[3], BH[4] and AH$_3$[5] at about 150 GPa to calculate $A_{X-H}$ with $b_{X-H} = A_{X-H} \cdot d_{X-H}$ as shown in Table S3.

Table S 3| The length of X-H bonds $b_{X-H}$ and its flexible factors $A_{X-H}$ suggested by model and calculated by DFT. The negative charge means that charge transfers from X. Here $b_{X-H} = A_{X-H} \cdot d_{X-H}$, $d_{X-H}$ is length of common X-H bonds in known binary hydrides.

|  | $b_{X-H}$, Å | $A_{X-H}$ | $b'_{X-H}$, Å | $A'_{X-H}$ | $d_{X-H}$, Å | transferred charge from X, electrons |
|---|---|---|---|---|---|---|
| LaBeH$_8$ | 1.25 | 91% | 1.36 | 99% | ~1.37 | -1.57 |
| LaBH$_8$ | 1.25 | 102% | 1.33 | 109% | ~1.22 | -1.2 |
| LaAlH$_8$ | 1.25 | 75% | 1.52 | 92% | ~1.65 | -2.28 |
| SrBH$_8$ | 1.44 | 118% | 1.32 | 108% | ~1.22 | -1.28 |
| CaBH$_8$ | 1.32 | 109% | 1.3 | 107% | ~1.22 | -1.29 |
| CaBeH$_8$ | 1.32 | 97% | 1.32 | 96% | ~1.37 | -1.51 |
| YBeH$_8$ | 1.19 | 86% | 1.34 | 98% | ~1.37 | -1.56 |

Compared with $A_{X-H}$ and $A'_{X-H}$, they change consistently relative to the common minimal lengths of X-H bond $d_{X-H}$, because geometric factor controls the length of X-H bonds. And our results give an idea for the replaceability of element X in AXH$_8$: elements, of which length of bonds is about 1.2~1.6 Å and electrons transfer to H, is possibly replaceable. Following this idea, we had found LaSiH$_8$ and LaPH$_8$, and LaSH$_8$ which are also potential high-Tc superconductors. According to (ii), although electronegativity of B is similar to H, but at pressure transferred charge from B to H is large, therefore, pressure could increase the electronegativity of H in this structure.

In summary, the structure of fluorite-type hydrides AXH$_8$ was described by the hard-sphere model, from which structural parameters were similar to those calculated by DFT at pressure about 100~200 GPa. Therefore, the novel structure of fluorite-type hydrides AXH$_8$ was proven to be stabilized by pressure. And the lengths of bonds in the XH$_8$ backbone were forced by the geometry of $AXH_8$. The length of X-H bond is about 1.2~1.6 Å. Because pressure increase the electronegativity in this structure, it is possible for H to get charge from elements with large electronegativity. Therefore, we had tried Si, P and S as candidates, which possess suitable bond lengths and electronegativity.

## 2. Computational details

High-pressure structural searches were performed from first principles using the AIRSS (*ab initio* Random Structure Searching) code[1], whose effectiveness has been confirmed by the successful applications to discovering the structures of solids, point defects, surfaces, and clusters. In pressure range of 50~200 GPa, we predict more than 15,000 structures in the La-Be-H system using variable composition structural searches, and the number of predicted structures at the $LaBeH_8$ stoichiometry is more than 1,000. And for systems Ca-Be-H, Ca-B-H, La-Al-H, La-B-H, Sr-B-H and Y-Be-H, each system we predicted around 5000 structures for their ternary convex hulls, and for each fluorite-type hydride we predicted around 1000 structures. Structural relaxations during the searches are performed using the CASTEP (Cambridge Sequential Total Energy Package) code with ultrasoft pseudopotentials[2]. Subsequent high-accuracy structural relaxations are performed using the projector augmented wave method (PAW) in the VASP code[3]. The cutoff energy was chosen to be 600 eV. The exchange-correlation functional was described using the Perdew-Burke-Ernzerhof (PBE) parametrization within the generalized gradient approximation (GGA)[4,5]. A Monkhorst-Pack k-point mesh of $2\pi \times 0.03$ Å$^{-1}$ was used to ensure that the enthalpy calculations were converged to within 1 meV/atom. We also perform Bader charge analysis to determine the extent of charge transfer[6,7]. Bonding in these hydrides was investigated by the crystal orbital Hamiltonian population (COHP)[8] analysis using LOBSTER code[9], which provides an atom-specific measure of the bonding character of states in a given energy region. We calculate electron-phonon matrix elements using density functional perturbation theory as implemented in the Quantum ESPRESSO package[10]. We use ultrasoft pseudopotentials[11] with an 80 Ry cutoff, an 18×18×18 *k*-point grid and a 6×6×6 *q*-point grid.

# 3. Equations for calculating Tc and related parameters

**(1) Self-consistent iteration solution of the Eliashberg equation**

The Migdal-Eliashberg equation has a form[12-14]:

$$\Delta(i\omega_n)Z(i\omega_n) = \frac{\pi T}{N_F}\sum_{n'}\frac{\Delta(i\omega_n')}{\sqrt{\omega_n'^2+\Delta^2(i\omega_n')}} \times [\lambda(\omega_n - \omega_{n'}) - N_F\mu^*]\delta(\epsilon) \qquad (S4)$$

$$Z(i\omega_n) = 1 + \frac{\pi T}{N_F\omega_n}\sum_{n'}\frac{\omega_n'}{\sqrt{\omega_n'^2+\Delta^2(i\omega_n')}}\lambda(\omega_n - \omega_{n'})\delta(\epsilon) \qquad (S5)$$

Here functions $\Delta(i\omega_n)$ and $Z(i\omega_n)$ represent pairing order parameter and the renormalization function, $N_F$ is the density of electronic states at the Fermi level, and $\delta(\epsilon)$ is the Dirac delta function. $i\omega_n = i(2n+1)\pi T_c$ are the fermion Matsubara frequencies (we employ the themodynamic Green's functions formalism; $\mu^*$ is the Coulomb pseudopotential, for which we use the widely accepted range of 0.1-0.13. $\lambda(\omega_n - \omega_{n'})$ contains the electron-phonon coupling matrix, phonon propagator, and the phonon density of states, and is given by:

$$\lambda(\omega_n - \omega_{n'}) = \int_0^\infty d\omega \frac{2\omega}{(\omega_n-\omega_n')^2+\omega^2}\alpha^2 F(\omega) \qquad (S6)$$

The equations for the order parameter and the renormalization function form a coupled nonlinear system and are solved self-consistently. We evaluated the renormalization function and the order parameter for each Matsubara frequency along the imaginary energy axis. After calculating $Z(i\omega_n)$ and $\Delta(i\omega_n)$, an analytic continuation is performed to the real axis using the Pade' functions. The calculation is performed for each T ($T_{min}<T \leq T_{max}$) ($T_{min} \approx 0$ and $T_{max} \geq T_c$). The critical temperature Tc is obtained as an asymptotic value as $\Delta(i\omega_n)$ tends to zero.

**(2) Gor'kov-Kresin's theory**

Gor'kov and Kresin (G-K) introduced the coupling constants $\lambda_{opt}$ and $\lambda_{ac}$ describing the interaction of electrons with optical and acoustic phonons[15,16]. The generalized Eliashberg equation has the form (at $T=T_c$):

$$\Delta(\omega_n)Z = \pi T \sum_{\omega_{n'}} \left[ \lambda_{\text{opt}} \frac{\widetilde{\Omega}_{\text{opt}}^2}{\widetilde{\Omega}_{\text{opt}}^2 + (\omega_n - \omega_{n'})^2} + \lambda_{\text{ac}} \frac{\widetilde{\Omega}_{\text{ac}}^2}{\widetilde{\Omega}_{\text{ac}}^2 + (\omega_n - \omega_{n'})^2} \right] \frac{\Delta(\omega_{n'})}{|\omega_{n'}|} \bigg|_{T=T_c}, \quad (S7)$$

$$\lambda_{\text{ac}} = 2 \int_0^{\omega_1} \frac{\alpha^2 F(\omega)}{\omega} d\omega, \quad \lambda_{\text{opt}} = 2 \int_{\omega_1}^{\omega_m} \frac{\alpha^2 F(\omega)}{\omega} d\omega, \quad \lambda_{\text{ac}} + \lambda_{\text{opt}} = \lambda, \quad (S8)$$

where $\omega_1$ is the maximum frequency for the acoustic modes, $\omega_m$ is the maximum frequency value. The mean square average frequency values are defined as follows:

$$\widetilde{\omega}_{\text{ac}} = \langle \omega_{\text{ac}}^2 \rangle^{\frac{1}{2}}, \quad \langle \omega_{\text{ac}}^2 \rangle = \frac{2}{\lambda_{\text{ac}}} \int_0^{\omega_1} d\omega \cdot \omega^2 \frac{\alpha^2 F(\omega)}{\omega} = \frac{2}{\lambda_{\text{ac}}} \int_0^{\omega_1} \alpha^2 F(\omega) \omega d\omega, \quad (S9)$$

$$\widetilde{\omega}_{\text{opt}} = \langle \omega_{\text{opt}}^2 \rangle^{\frac{1}{2}}, \quad \langle \omega_{\text{opt}}^2 \rangle = \frac{2}{\lambda_{\text{opt}}} \int_{\omega_1}^{\omega_m} d\omega \cdot \omega^2 \frac{\alpha^2 F(\omega)}{\omega} = \frac{2}{\lambda_{\text{opt}}} \int_{\omega_1}^{\omega_m} \alpha^2 F(\omega) \omega d\omega, \quad (S10)$$

For our predicted hydrides the $\lambda_{\text{ac}} \ll \lambda_{\text{opt}}$, we assume that:

$$T_c = T_c^{opt} + \Delta T_c^{ac}, \text{ and } T_c^{opt} \gg \Delta T_c^{ac} \quad (S11)$$

As a result, the expression for Tc can be written in the form:

$$T_c = \left[ 1 + 2 \frac{\lambda_{\text{ac}}}{\lambda_{\text{opt}} - \mu^*} \cdot \frac{1}{1+\rho^{-2}} \right] T_c^0, \rho = \frac{\widetilde{\omega}_{\text{ac}}}{\pi T_c^0}, T_c^0 \equiv T_c^{opt}. \quad (S12)$$

Here the $T_c^0$ is defined as the transition temperatures caused by the interaction of electrons with optical phonons only; for $\lambda_{\text{opt}} \leq 1.5$:

$$T_c^0 = \frac{\widetilde{\omega}_{\text{opt}}}{1.2} \exp \left[ -\frac{1.04(1+\lambda_{\text{opt}})}{\lambda_{\text{opt}} - \mu^*(1+0.62\lambda_{\text{opt}})} \right]. \quad (S13)$$

For $\lambda_{\text{opt}} > 1.5$:

$$T_c^0 = \frac{0.25 \widetilde{\omega}_{\text{opt}}}{[e^{\frac{2}{\lambda_{eff}}} - 1]^{1/2}}. \quad (S14)$$

Here the $\lambda_{\text{eff}}$ is defined as follows:

$$\lambda_{\text{eff}} = (\lambda_{\text{opt}} - \mu^*)[1 + 2\mu^* + \lambda_{\text{opt}} \mu^* t(\lambda_{\text{opt}})]^{-1}, \quad (S15)$$

$$t(x) = 1.5 \exp(-0.28) x. \quad (S16)$$

We used the following expression:

$$\alpha = \frac{1}{2}\left[1 - 4\frac{\lambda_{ac}}{\lambda_{opt}}\frac{\rho^2}{(\rho^2+1)^2}\right] \qquad (S17)$$

to calculate the isotope coefficient α.

### (3) Allen-Dynes modified McMillan eqation

The Allen-Dynes modified McMillan equation which is the approximate analytic solution of $T_c$ from Eliashberg equations at $\omega \approx 0$ or $\infty$ and $\lambda \ll 1.5$[17]:

$$T_c = \frac{\omega_{log}}{1.2}\exp\left[-\frac{1.04(1+\lambda)}{\lambda-\mu^*(1+0.62\lambda)}\right], \qquad (S18)$$

when $\lambda > 1.5$, we have the correction equation as follows:

$$T_c = \frac{f_1 f_2 \omega_{log}}{1.2}\exp\left[-\frac{1.04(1+\lambda)}{\lambda-\mu^*(1+0.62\lambda)}\right], \qquad (S19)$$

two separate correction factors $f_1$ and $f_2$ are given by:

$$f_1 = \sqrt[3]{\left[1+\left(\frac{\lambda}{2.46(1+3.8\mu^*)}\right)^{\frac{3}{2}}\right]}, \quad f_2 = 1 + \frac{\left(\frac{\bar{\omega}_2}{\omega_{log}}-1\right)\lambda^2}{\lambda^2+[1.82(1+6.3\mu^*)\frac{\bar{\omega}_2}{\omega_{log}}]^2}, \qquad (S20)$$

$$\bar{\omega}_2 = \sqrt{\frac{2}{\lambda}\int \alpha^2 F(\omega)\,\omega d\omega}, \qquad (S21)$$

here $\bar{\omega}_2$ is mean square frequency, $\omega_{log}$ is the logarithmic average frequency and $\mu^*$ is the Coulomb pseudopotential, for which we use the widely accepted range of 0.1-0.13. The $\omega_{log}$ and EPC constant λ were calculated as:

$$\omega_{log} = \exp\left[\frac{2}{\lambda}\int \frac{d\omega}{\omega}\alpha^2 F(\omega)\ln\omega\right], \qquad (S22)$$

$$\lambda = 2\int \frac{\alpha^2 F(\omega)}{\omega}d\omega = \sum_{qj}\lambda_{qj}w(q), \qquad (S23)$$

$$\lambda_{qj} = \frac{\gamma_{qj}}{\pi\hbar N(\varepsilon_f)\omega_{qj}^2}, \qquad (S24)$$

where $\lambda_{qj}$ is the mode EPC parameter and $w(q)$ is the weight of phonon mode $j$ at wave vector $q$ in the first Brillouin zone (BZ).

Table S4. The calculated EPC parameter λ, logarithmic average phonon frequency ω log (K), electronic density of states at Fermi level N($\varepsilon_f$) (states/spin/Ry/f.u.) and superconducting transition

temperatures $T_c$ (K) of flouride-type hydrides with $\mu^*=0.1$ at corresponding pressures (GPa). And the values of $T_c$ are calculated from Elashberg equation with use of the approach with many iteration (IA), and with use of the Gorkov-Kresin theory (G-K). The table contains also the values obtained from the Allen- Dynes modified McMillan equation (Mc-A-D, See Eq. S19).

| Hydrides | Pressure | $\lambda$ | $\omega_{log}$ | $N(\varepsilon_f)$ | $T_c$ (Mc-A-D) | $T_c$ (IA) | $\lambda_{ac}$ | $\lambda_{opt}$ | $T_c$ (G-K) |
|---|---|---|---|---|---|---|---|---|---|
| LaBeH$_8$ | 50 | 2.19 | 897.63 | 6.94 | 167 | 191 | 0.25 | 1.94 | 165 |
| LaBH$_8$ | 70 | 1.98 | 843.63 | 4.63 | 144 | 160 | 0.30 | 1.67 | 155 |
| LaAlH$_8$ | 100 | 1.81 | 839.68 | 5.21 | 130 | 144 | 0.17 | 1.58 | 140 |
| CaBeH$_8$ | 210 | 3.91 | 798.06 | 5.02 | 254 | 302 | 1.12 | 2.77 | 291 |
| CaBH$_8$ | 100 | 3.44 | 767.67 | 4.50 | 212 | 238 | 0.84 | 2.49 | 237 |
| YBeH$_8$ | 100 | 3.05 | 855.70 | 5.67 | 215 | 249 | 0.45 | 2.60 | 226 |
| SrBH$_8$ | 150 | 1.79 | 1039.12 | 4.62 | 163 | 200 | 0.32 | 1.47 | 177 |
| LaSH$_8$ | 200 | 2.37 | 827.19 | 8.42 | 169 | 195 | 0.44 | 1.92 | 190 |
| LaSiH$_8$ | 100 | 2.41 | 659.62 | 5.17 | 137 | 150 | 0.49 | 1.73 | 159 |
| LaPH$_8$ | 200 | 1.41 | 1090.43 | 7.05 | 137 | 151 | 0.12 | 1.35 | 142 |

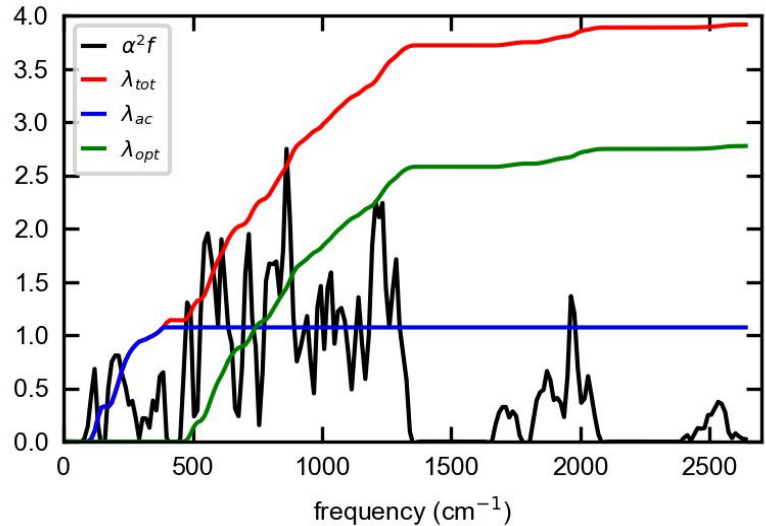

**FIG. S4.** Calculated Eliashberg spectral function $\alpha^2F(\omega)$ (black line), total electron-phonon integral $\lambda_{tot}$ (red line), electron and acoustic phonon integral $\lambda_{ac}$ (blue line) and electron and optical phonon integral $\lambda_{opt}$ (green line) of alloyed ternary hydrides CaBeH$_8$.

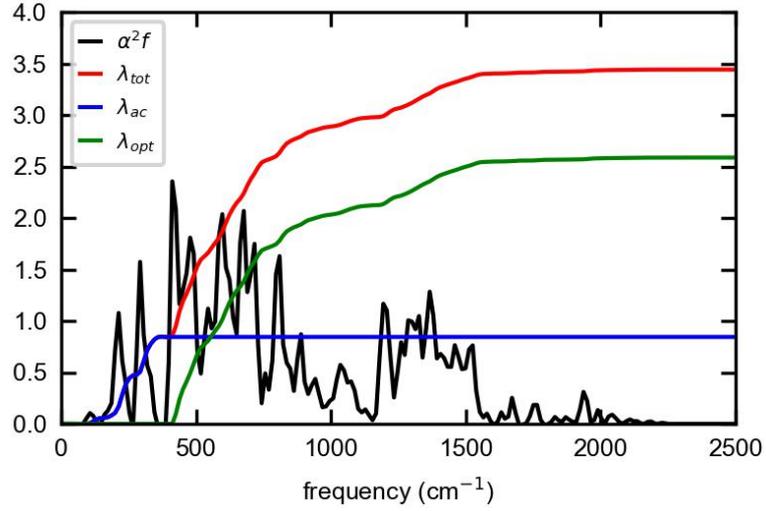

**FIG. S5.** Calculated Eliashberg spectral function $\alpha^2F(\omega)$ (black line), total electron-phonon integral $\lambda_{tot}$ (red line), electron and acoustic phonon integral $\lambda_{ac}$ (blue line) and electron and optical phonon integral $\lambda_{opt}$ (green line) of alloyed ternary hydrides $CaBH_8$.

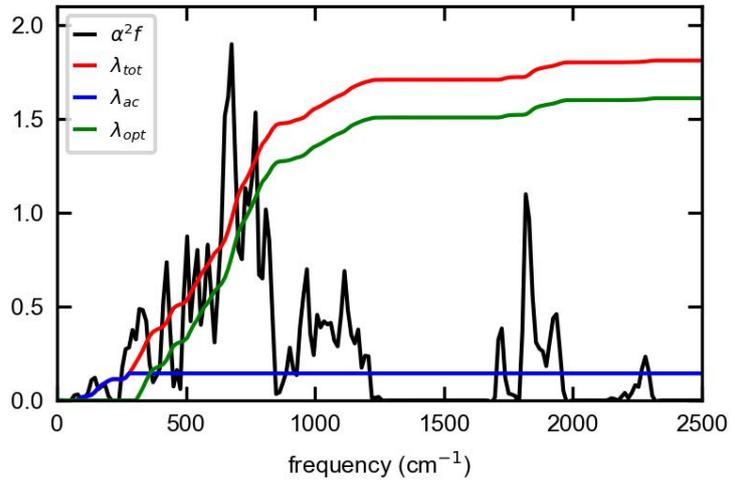

**FIG. S6.** Calculated Eliashberg spectral function $\alpha^2F(\omega)$ (black line), total electron-phonon integral $\lambda_{tot}$ (red line), electron and acoustic phonon integral $\lambda_{ac}$ (blue line) and electron and optical phonon integral $\lambda_{opt}$ (green line) of alloyed ternary hydrides $LaAlH_8$.

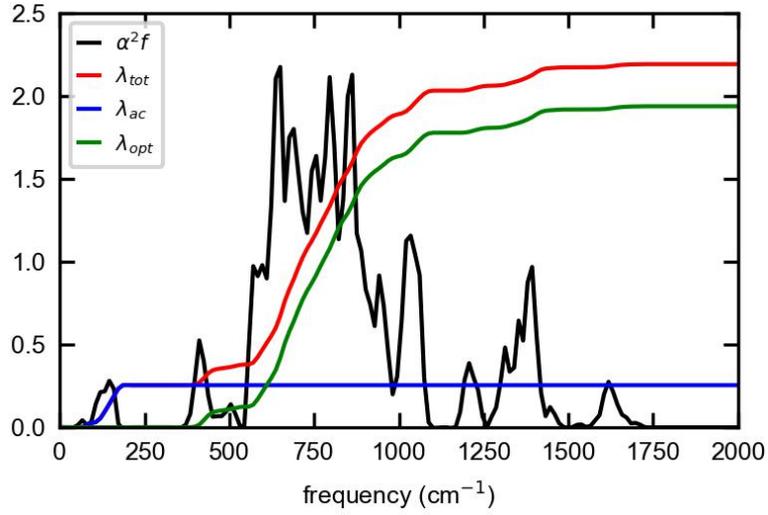

**FIG. S7.** Calculated Eliashberg spectral function $\alpha^2F(\omega)$ (black line), total electron-phonon integral $\lambda_{tot}$ (red line), electron and acoustic phonon integral $\lambda_{ac}$ (blue line) and electron and optical phonon integral $\lambda_{opt}$ (green line) of alloyed ternary hydrides $LaBeH_8$.

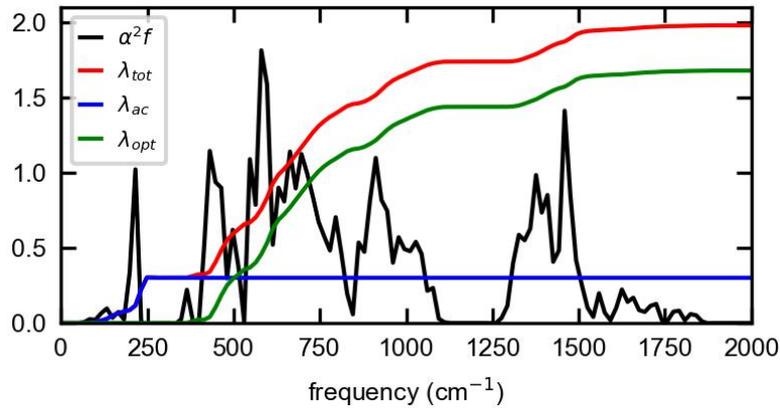

**FIG. S8.** Calculated Eliashberg spectral function $\alpha^2F(\omega)$ (black line), total electron-phonon integral $\lambda_{tot}$ (red line), electron and acoustic phonon integral $\lambda_{ac}$ (blue line) and electron and optical phonon integral $\lambda_{opt}$ (green line) of alloyed ternary hydrides $LaBH_8$.

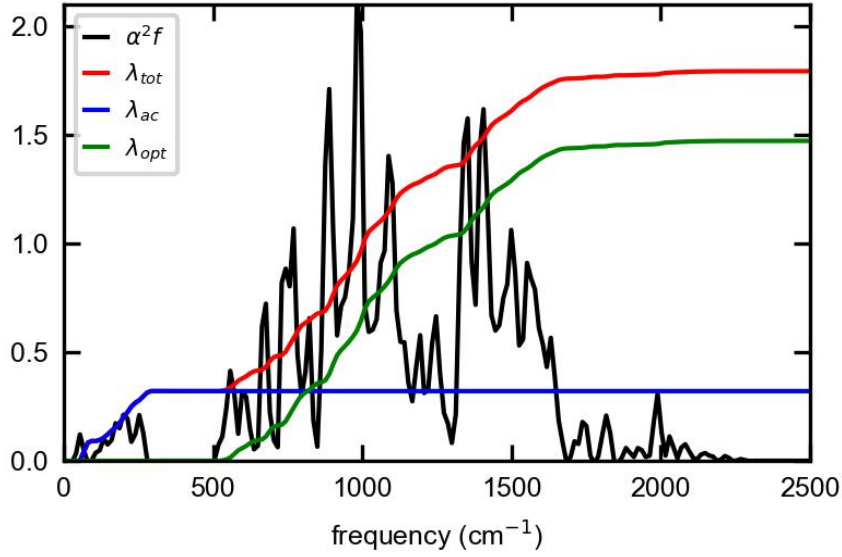

**FIG. S9.** Calculated Eliashberg spectral function $\alpha^2F(\omega)$ (black line), total electron-phonon integral $\lambda_{tot}$ (red line), electron and acoustic phonon integral $\lambda_{ac}$ (blue line) and electron and optical phonon integral $\lambda_{opt}$ (green line) of alloyed ternary hydrides SrBH$_8$.

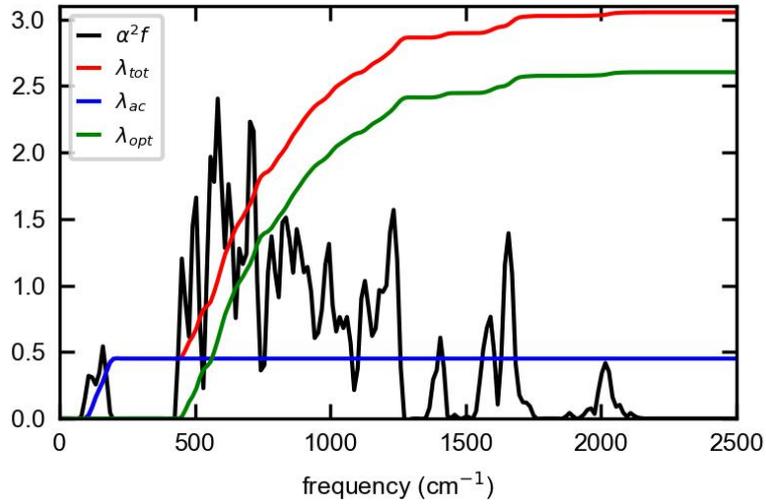

**FIG. S10.** Calculated Eliashberg spectral function $\alpha^2F(\omega)$ (black line), total electron-phonon integral $\lambda_{tot}$ (red line), electron and acoustic phonon integral $\lambda_{ac}$ (blue line) and electron and optical phonon integral $\lambda_{opt}$ (green line) of alloyed ternary hydrides YBeH$_8$.

# 4. Structural information

| Space group Pressure | Lattice parameters (Å) | Atomic coordinates (fractional) | | | | Sites |
|---|---|---|---|---|---|---|
| $Fm\bar{3}m$-LaBeH$_8$ 50 GPa | a = b = c = 5.643 | H | 0.351 | -0.351 | -0.351 | 32f |
| | | Be | 0.500 | 0.500 | 0.500 | 4b |
| | | La | 0.000 | 0.000 | 0.000 | 4a |
| $Fm\bar{3}m$-LaBH$_8$ 70 GPa | a = b = c = 5.456 | H | -0.647 | 0.352 | -0.647 | 32f |
| | | B | 0.500 | 0.500 | 0.500 | 4b |
| | | La | 0.000 | 0.000 | 0.000 | 4a |
| $Fm\bar{3}m$-LaAlH$_8$ 100 GPa | a = b = c = 5.582 | H | 0.338 | -0.338 | 0.162 | 32f |
| | | B | 0.500 | 0.500 | 0.500 | 4b |
| | | La | 0.000 | 0.000 | 0.000 | 4a |
| $Fm\bar{3}m$-LaSiH$_8$ 100 GPa | a = b = c = 5.540 | H | -0.340 | 0.660 | 0.660 | 32f |
| | | Si | 0.500 | 0.500 | 0.500 | 4b |
| | | La | 0.000 | 0.000 | 0.000 | 4a |
| $Fm\bar{3}m$-LaSH$_8$ 200 GPa | a = b = c = 5.173 | H | -0.334 | 0.666 | 0.666 | 32f |
| | | S | 0.500 | 0.500 | 0.500 | 4b |
| | | La | 0.000 | 0.000 | 0.000 | 4a |
| $Fm\bar{3}m$-LaPH$_8$ 200 GPa | a = b = c = 5.186 | H | -0.337 | 0.663 | 0.663 | 32f |
| | | P | 0.500 | 0.500 | 0.500 | 4b |
| | | La | 0.000 | 0.000 | 0.000 | 4a |
| $Fm\bar{3}m$-CaBH$_8$ 200 GPa | a = b = c = 5.074 | H | -0.653 | 0.347 | 0.347 | 32f |
| | | B | 0.500 | 0.500 | 0.500 | 4b |
| | | Ca | 0.000 | 0.000 | 0.000 | 4a |
| $Fm\bar{3}m$-CaBeH$_8$ | a = b = c = 4.752 | H | 0.344 | -0.344 | 0.344 | 32f |

| | | | | | |
|---|---|---|---|---|---|
| 210 GPa | | Be | 0.500 | 0.500 | 0.500 | 4b |
| | | Ca | 0.000 | 0.000 | 0.000 | 4a |
| $Fm\bar{3}m$-SrBH$_8$ 150 GPa | a = b = c = 5.051 | H | -0.349 | 0.349 | -0.151 | 32f |
| | | B | 0.500 | 0.500 | 0.500 | 4b |
| | | Sr | 0.000 | 0.000 | 0.000 | 4a |
| $Fm\bar{3}m$-YBeH$_8$ 100 GPa | a = b = c = 5.199 | H | -0.846 | 0.654 | -0.654 | 32f |
| | | Be | 0.500 | 0.500 | 0.500 | 4b |
| | | Y | 0.000 | 0.000 | 0.000 | 4a |
| $F\bar{4}3m$-LaBeH$_5$ 50 GPa | a = b = c = 5.027 | H1 | 0.606 | -0.606 | -0.394 | 16e |
| | | H2 | 0.000 | 0.000 | 0.000 | 4a |
| | | Be | 0.250 | 0.750 | 0.750 | 4c |
| | | La | 0.750 | 0.250 | 0.250 | 4b |
| $P\bar{1}$-LaBeH$_8$ 100 GPa | a = 3.394 | H1 | 0.157 | 0.157 | 0.487 | 1a |
| | b = 5.130 | H2 | 0.071 | 0.874 | 0.131 | 1a |
| | c = 5.219 | H3 | 0.843 | 0.844 | 0.513 | 1a |
| | α = 73.0 | H4 | 0.929 | 0.126 | 0.869 | 1a |
| | β = 71.2 | H5 | 0.341 | 0.934 | 0.338 | 1a |
| | γ = 74.2 | H6 | 0.659 | 0.066 | 0.662 | 1a |
| | | H7 | 0.329 | 0.319 | 0.897 | 1a |
| | | H8 | 0.671 | 0.680 | 0.103 | 1a |
| | | H9 | 0.496 | 0.214 | 0.692 | 1a |
| | | H10 | 0.504 | 0.786 | 0.308 | 1a |
| | | H11 | 0.811 | 0.515 | 0.845 | 1a |
| | | H12 | 0.188 | 0.485 | 0.155 | 1a |
| | | H13 | 0.711 | 0.490 | 0.510 | 1a |
| | | H14 | 0.289 | 0.509 | 0.490 | 1a |

| | | | | |
|---|---|---|---|---|
| H15 | 0.414 | 0.077 | 0.008 | 1a |
| H16 | 0.586 | 0.923 | 0.992 | 1a |
| Be1 | 0.001 | 0.656 | 0.336 | 1a |
| Be2 | 0.999 | 0.344 | 0.663 | 1a |
| La1 | 0.255 | 0.759 | 0.769 | 1a |
| La2 | 0.745 | 0.241 | 0.231 | 1a |

# 5. The ternary convex hull of "fluorite-like" backbone hydride

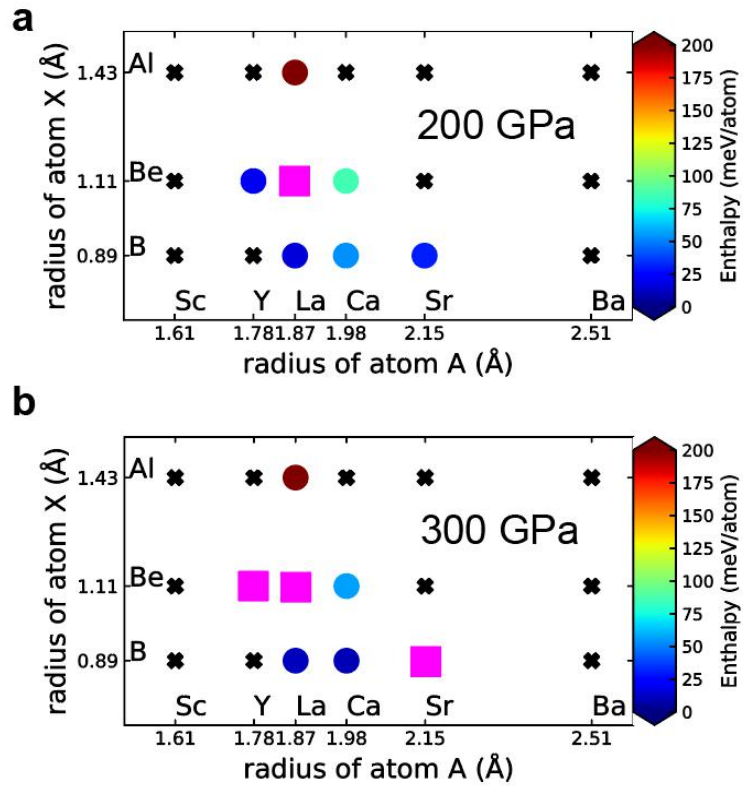

**FIG. S11.** Calculated enthalpy of alloyed ternary hydrides $AXH_8$ above the convex hull at (a) 200 GPa and (b) 300 GPa. The radius of atom A is plotted on the x-axis and the radius of atom X on the y-axis. Dynamically unstable systems are shown as black cross marks. Metastable phases are shown as circles, colored according to the calculated enthalpy above the convex hull. Thermodynamically stable phases are shown as carmine squares.

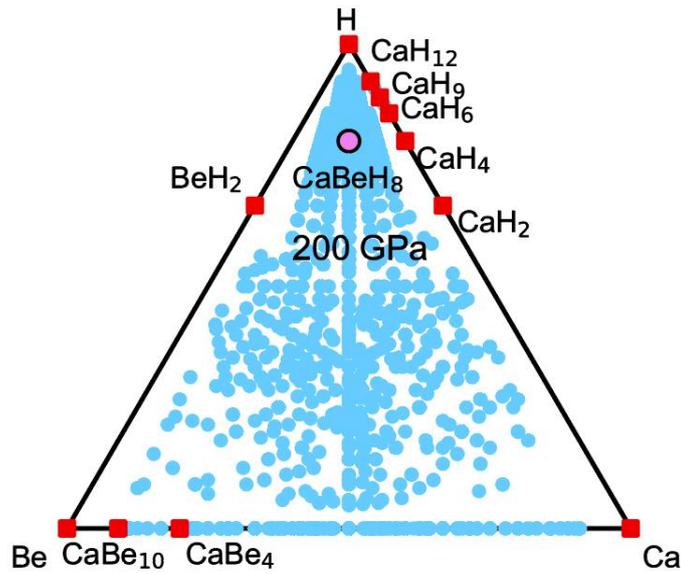

**FIG. S12.** The ternary phase diagram of Ca, Be, and H at 200 GPa. The corresponding elements and boundary binary phases are chosen from the results of the previous works. Light blue circles indicate metastable phases. Red squares indicate stable phases. And the big purple circle with edge indicates component $CaBeH_8$, which is a metastable phase. The structures of binary hydrides were from ref. [18-20].

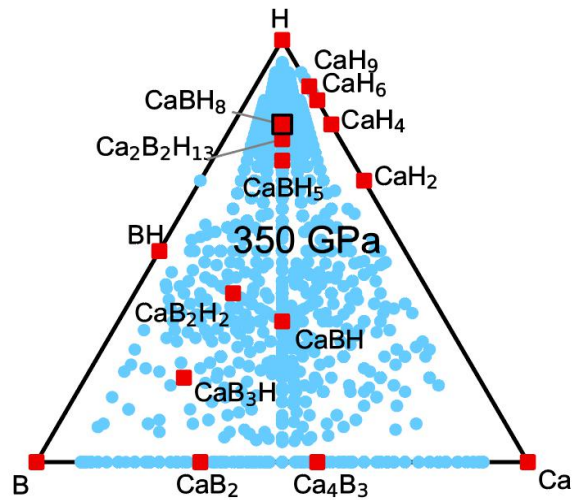

**FIG. S13.** The ternary phase diagram of Ca, B, and H at 350 GPa. The corresponding elements and boundary binary phases are chosen from the results of the previous works. Light blue circles indicate metastable phases. Red squares indicate stable phases. And the big red square with edge indicates component $CaBH_8$, which is a stable phase. The structures of binary hydrides were from ref. [19-21].

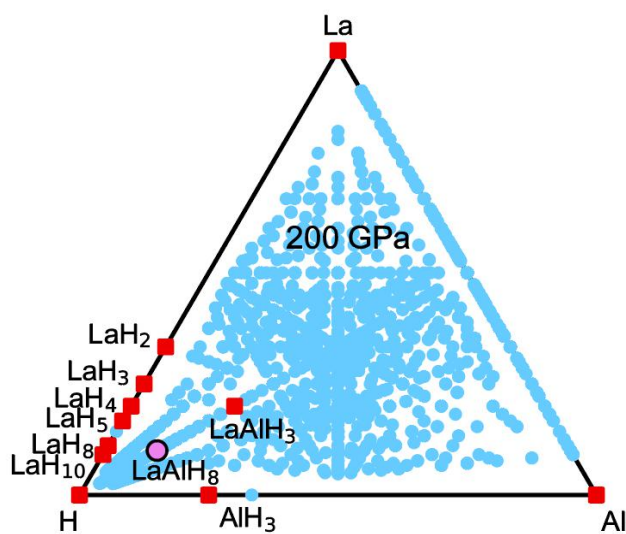

**FIG. S14.** The ternary phase diagram of La, Al, and H at 200 GPa. The corresponding elements and boundary binary phases are chosen from the results of the previous works. Light blue circles indicate metastable phases. Red squares indicate stable phases. And the big purple circle with edge indicates component LaAlH$_8$, which is a metastable phase. The structures of binary hydrides were from ref. [22,23].

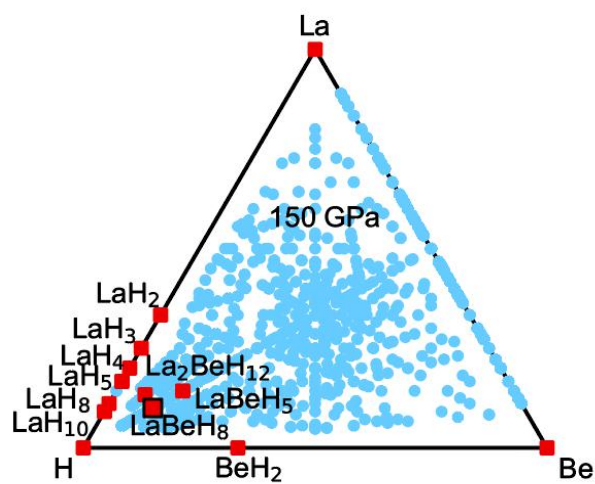

**FIG. S15.** The ternary phase diagram of La, Be, and H at 150 GPa. The corresponding elements and boundary binary phases are chosen from the results of the previous works. Light blue circles indicate metastable phases. Red squares indicate stable phases. And the big purple circle with edge indicates component LaBeH$_8$, which is a stable phase. The structures of binary hydrides were from ref. [18,23].

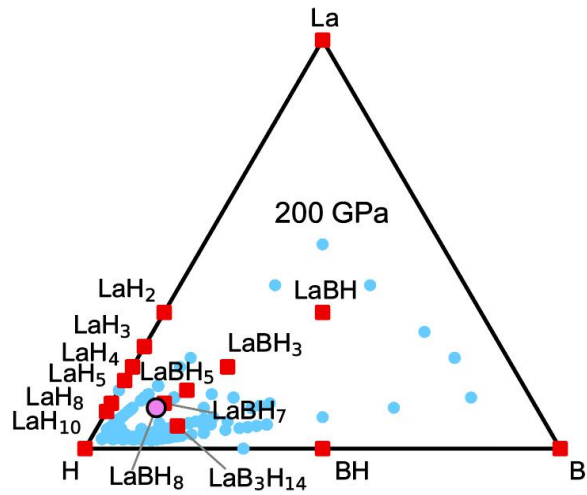

**FIG. S16.** The ternary phase diagram of La, B, and H at 200 GPa. The corresponding elements and boundary binary phases are chosen from the results of the previous works. Light blue circles indicate metastable phases. Red squares indicate stable phases. And the big purple circle with edge indicates component $LaBH_8$, which is a dynamically stable phase. The structures of binary hydrides were from ref. [21,23].

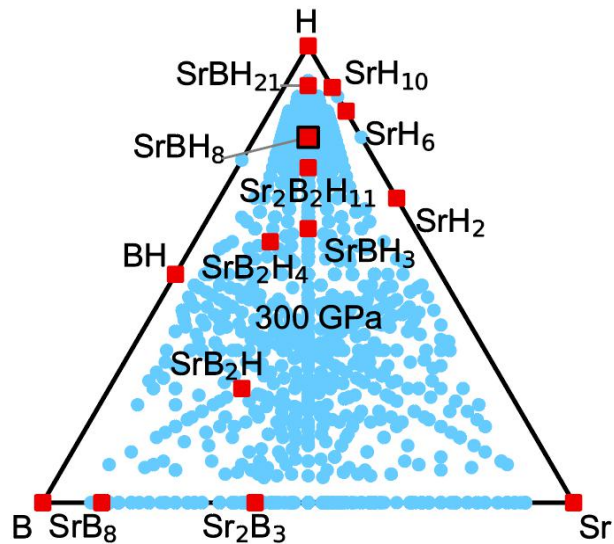

**FIG. S17.** The ternary phase diagram of Sr, B, and H at 300 GPa. The corresponding elements and boundary binary phases are chosen from the results of the previous works. Light blue circles indicate metastable phases. Red squares indicate stable phases. And the big purple circle with edge indicates component $SrBH_8$, which is a stable phase. The structures of binary hydrides were from ref. [21,24].

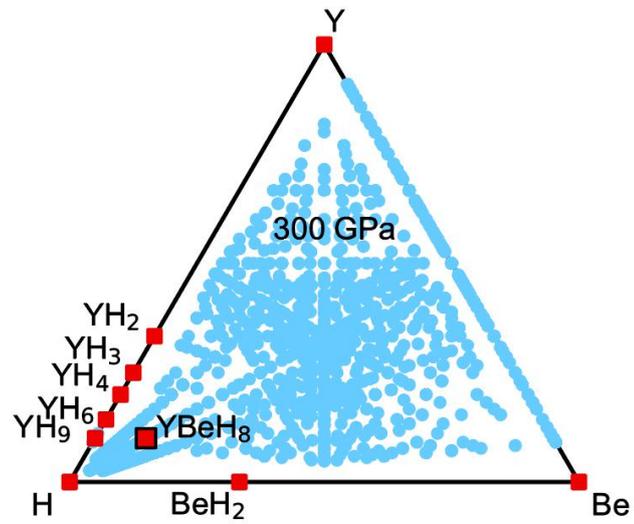

**FIG. S18.** The ternary phase diagram of Y, Be, and H at 300 GPa. The corresponding elements and boundary binary phases are chosen from the results of the previous works. Light blue circles indicate metastable phases. Red squares indicate stable phases. And the big red square with edge indicates component YBeH$_8$, which is a stable phase. The structures of binary hydrides were from ref. [18,23].

## 6. The Crystalline Orbital Hamiltonian Population (COHP) and Integrated Orbital Hamiltonian Population (ICOHP) of "fluorite-like" backbone hydride

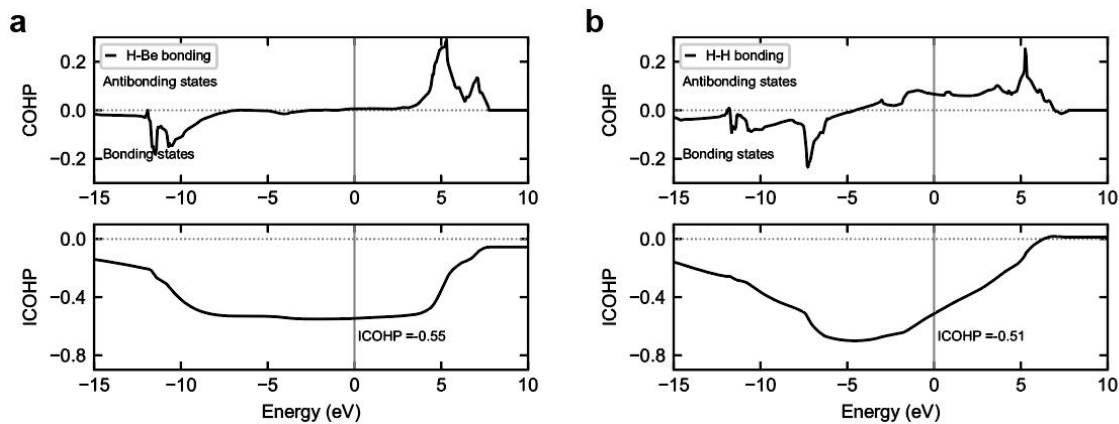

**FIG. S19.** The calculated Crystalline Orbital Hamiltonian Population (COHP, top panel) and Integrated Crystalline Orbital Hamiltonian Population (ICOHP, bottom panel) of **a** H-Be bonds **b** H-H bonds of CaBeH$_8$ at 200 GPa.

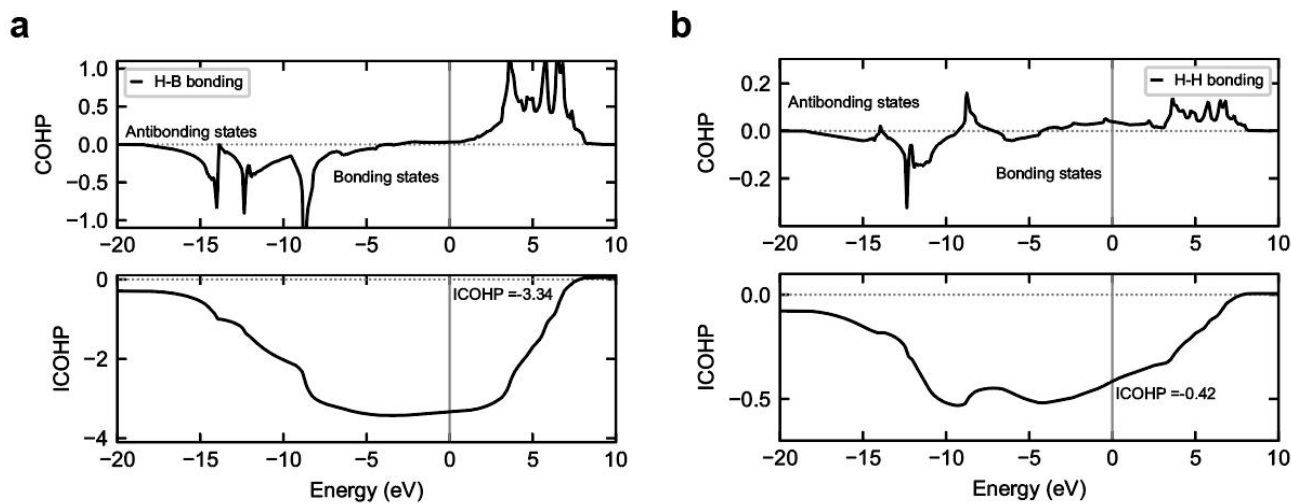

**FIG. S20.** The calculated Crystalline Orbital Hamiltonian Population (COHP, top panel) and Integrated Crystalline Orbital Hamiltonian Population (ICOHP, bottom panel) of **a** H-B bonds **b** H-H bonds of CaBH$_8$ at 100 GPa.

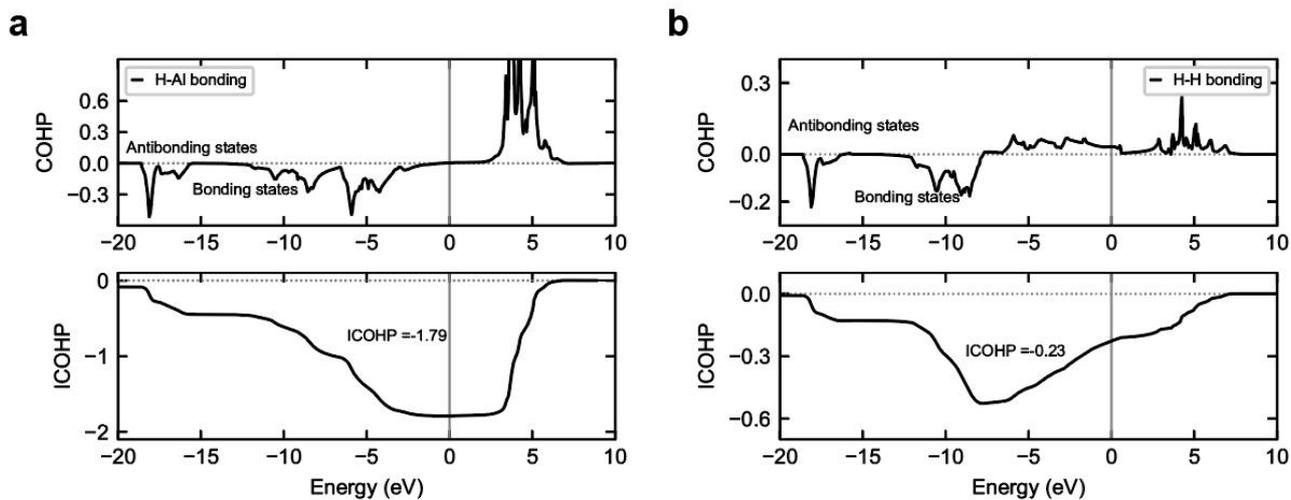

**FIG. S21.** The calculated Crystalline Orbital Hamiltonian Population (COHP, top panel) and Integrated Crystalline Orbital Hamiltonian Population (ICOHP, bottom panel) of **a** H-Al bonds **b** H-H bonds of LaAlH$_8$ at 100 GPa.

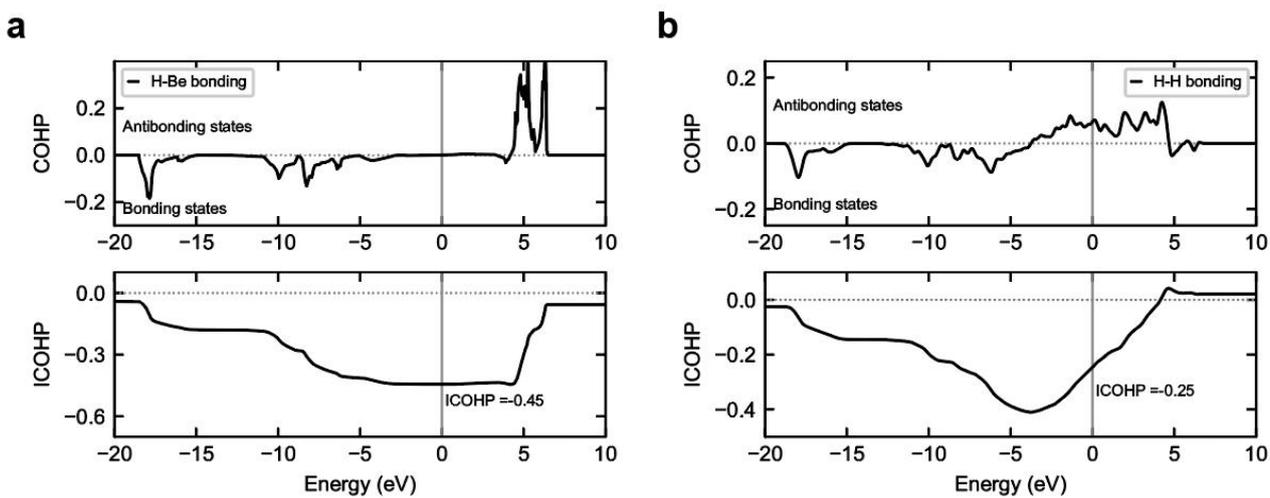

**FIG. S22.** The calculated Crystalline Orbital Hamiltonian Population (COHP, top panel) and Integrated Crystalline Orbital Hamiltonian Population (ICOHP, bottom panel) of **a** H-Be bonds **b** H-H bonds of LaBeH$_8$ at 100 GPa.

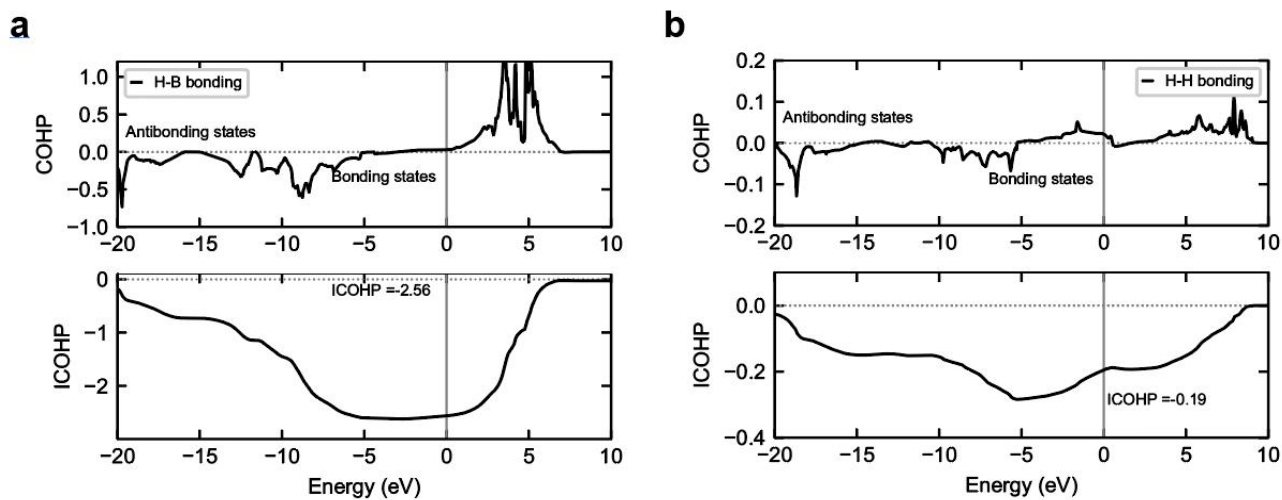

**FIG. S23.** The calculated Crystalline Orbital Hamiltonian Population (COHP, top panel) and Integrated Crystalline Orbital Hamiltonian Population (ICOHP, bottom panel) of **a** H-B bonds **b** H-H bonds of LaBH$_8$ at 100 GPa.

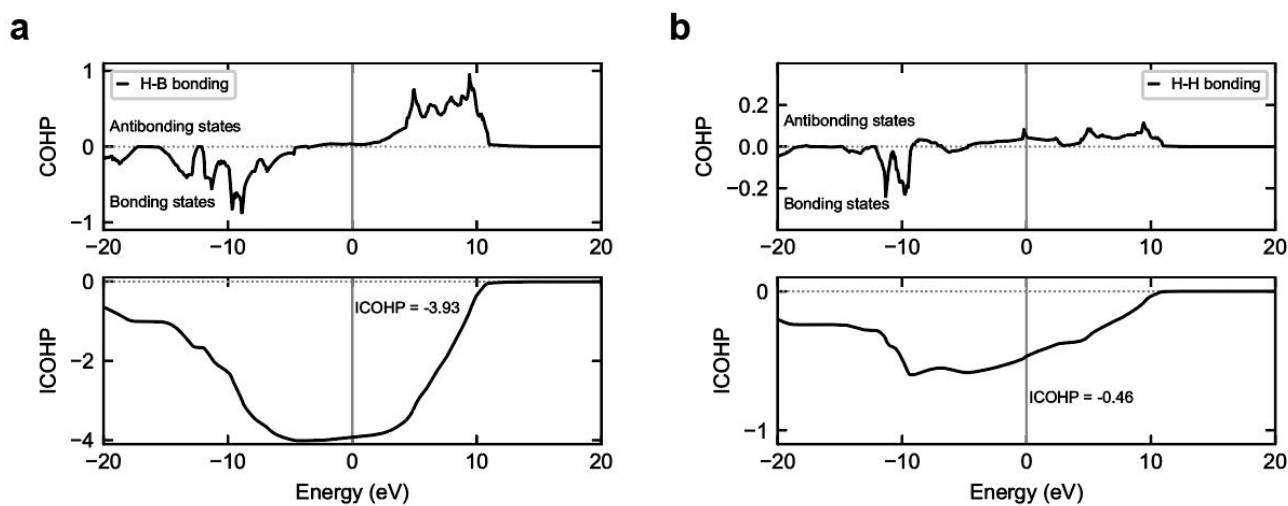

**FIG. S24.** The calculated Crystalline Orbital Hamiltonian Population (COHP, top panel) and Integrated Crystalline Orbital Hamiltonian Population (ICOHP, bottom panel) of **a** H-B bonds **b** H-H bonds of SrBH$_8$ at 200 GPa.

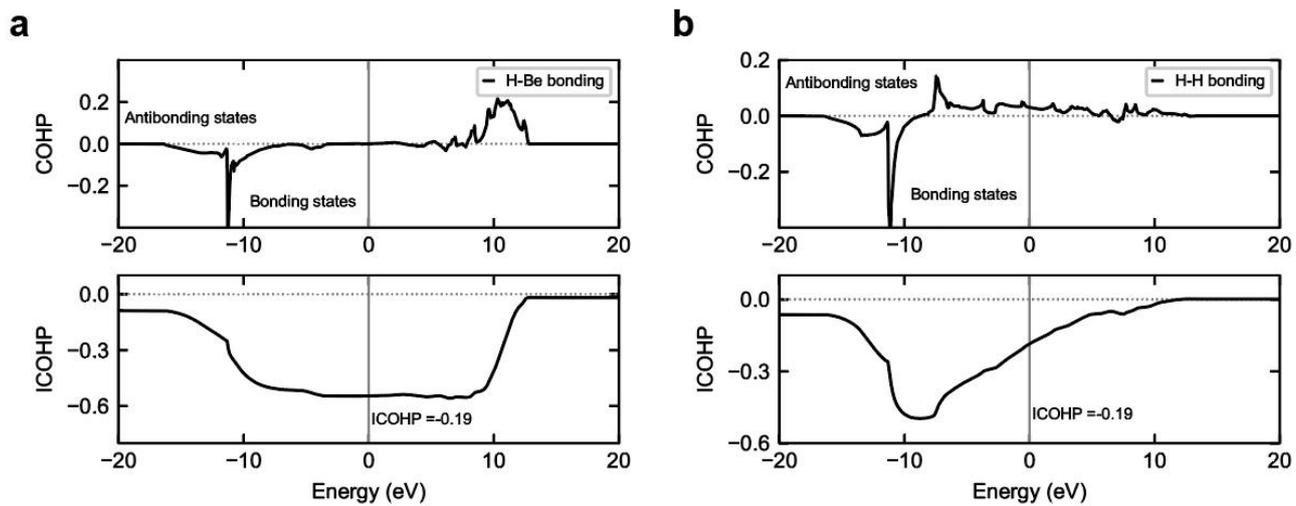

**FIG. S25.** The calculated Crystalline Orbital Hamiltonian Population (COHP, top panel) and Integrated Crystalline Orbital Hamiltonian Population (ICOHP, bottom panel) of **a** H-Be bonds **b** H-H bonds of YBeH$_8$ at 100 GPa.

# 7. Electronic structure of "fluorite-like" backbone hydride

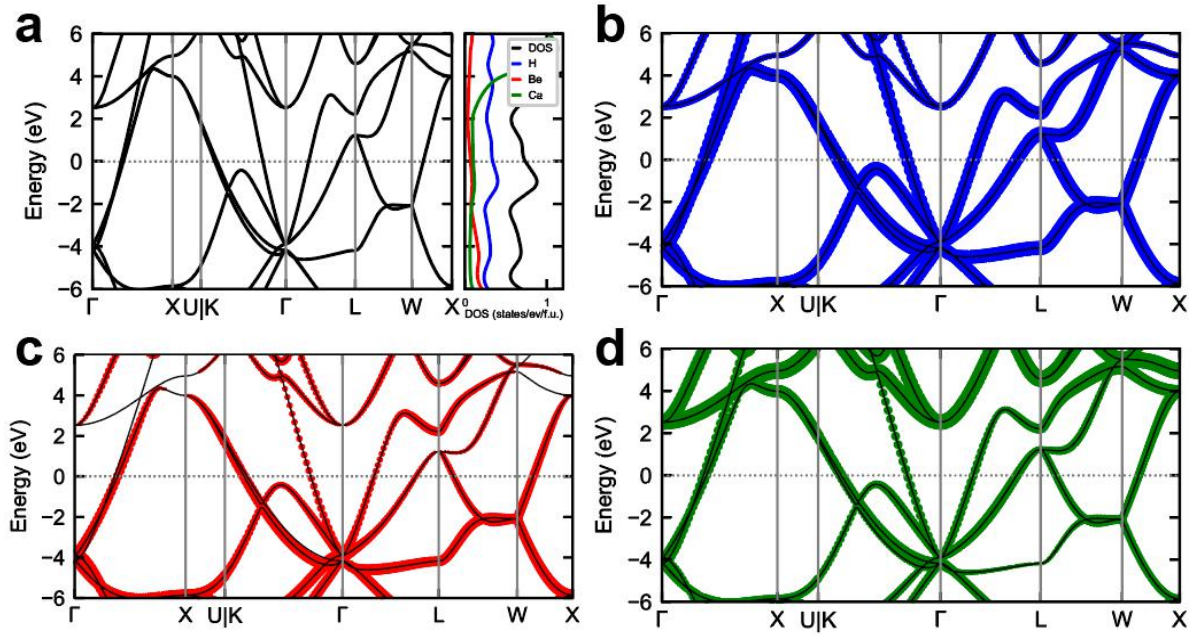

**FIG. S26.** Electronic structure of The of CaBeH$_8$ at 200 GPa. **a** Electronic band structure (left panel) and projected density of states (right panel). And fatband of **b** H, **c** Be and **d** Ca.

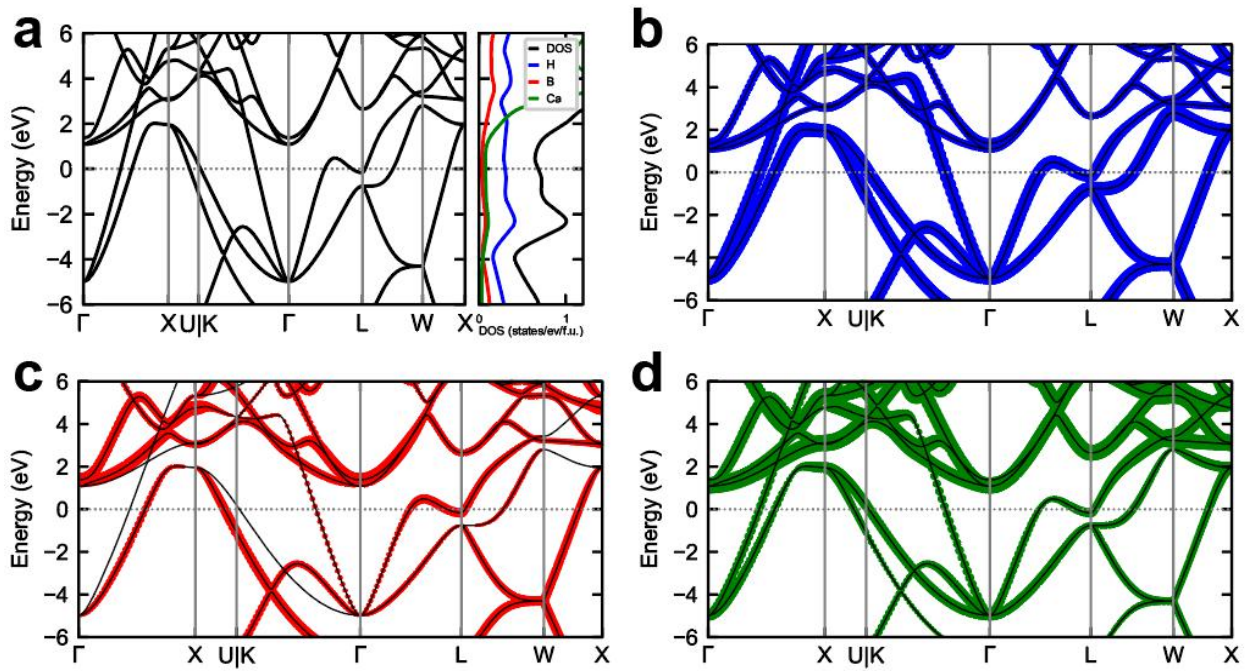

**FIG. S27.** Electronic structure of The of CaBH$_8$ at 100 GPa. **a** Electronic band structure (left panel) and projected density of states (right panel). And fatband of **b** H, **c** B and **d** Ca.

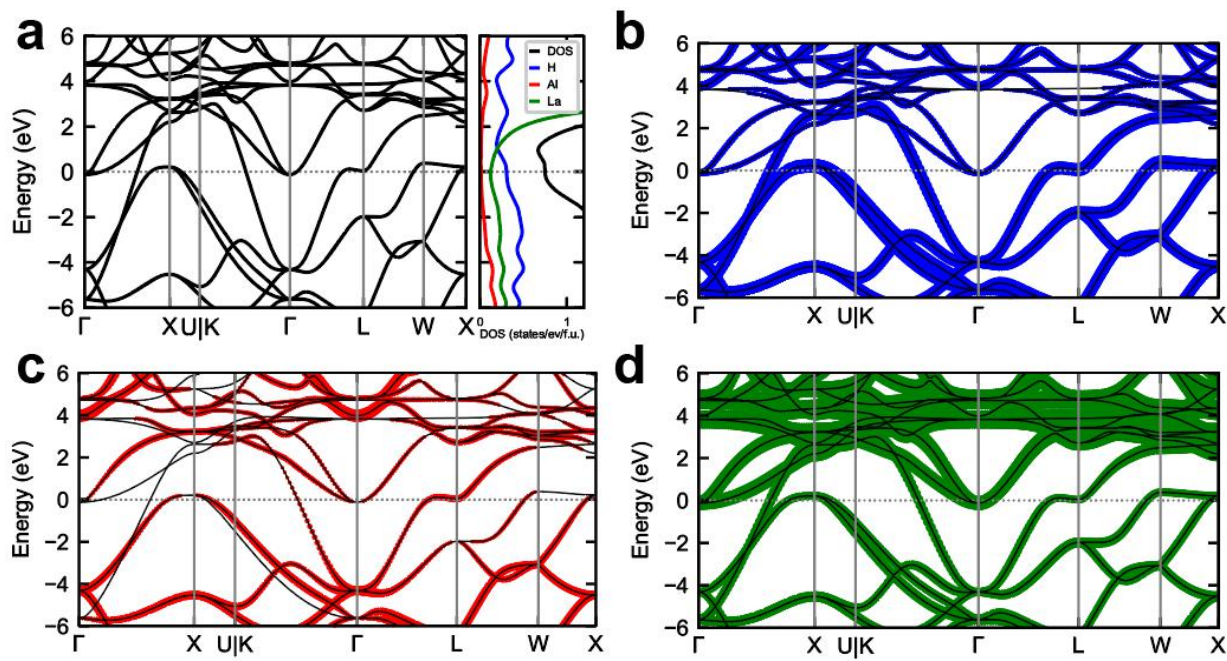

**FIG. S28.** Electronic structure of The of LaAlH$_8$ at 100 GPa. **a** Electronic band structure (left panel) and projected density of states (right panel). And fatband of **b** H, **c** Al and **d** La.

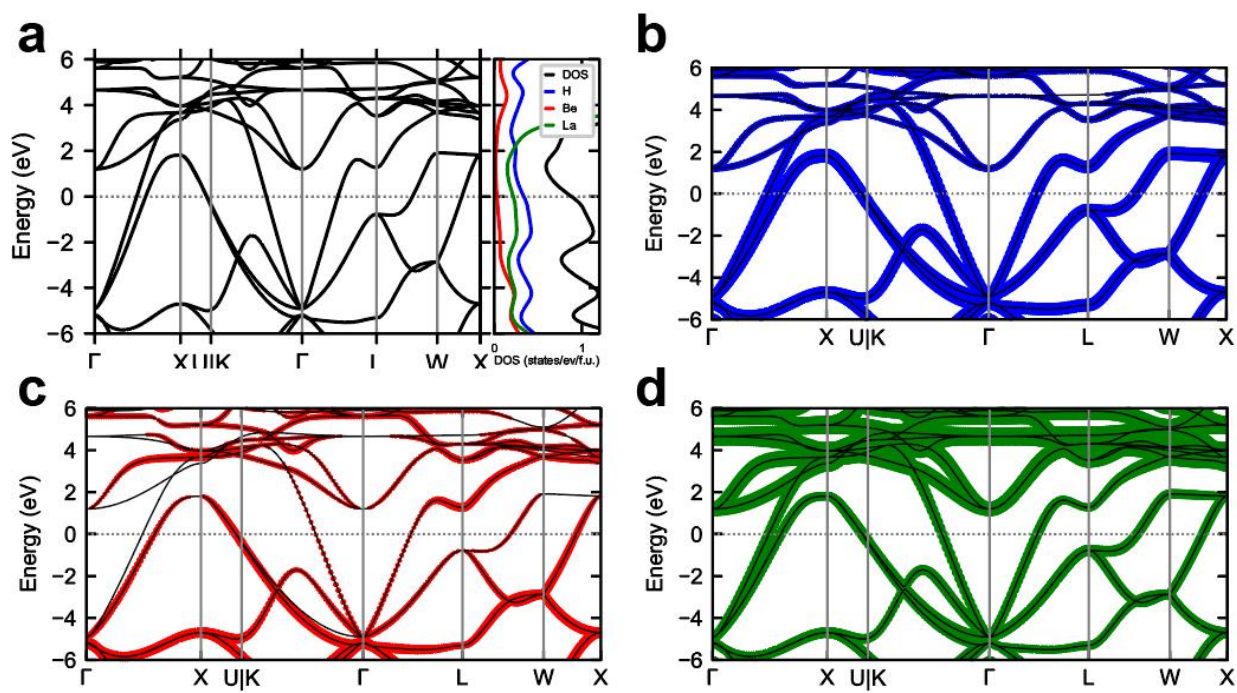

**FIG. S29.** Electronic structure of The of LaBeH$_8$ at 100 GPa. **a** Electronic band structure (left panel) and projected density of states (right panel). And fatband of **b** H, **c** Be and **d** La.

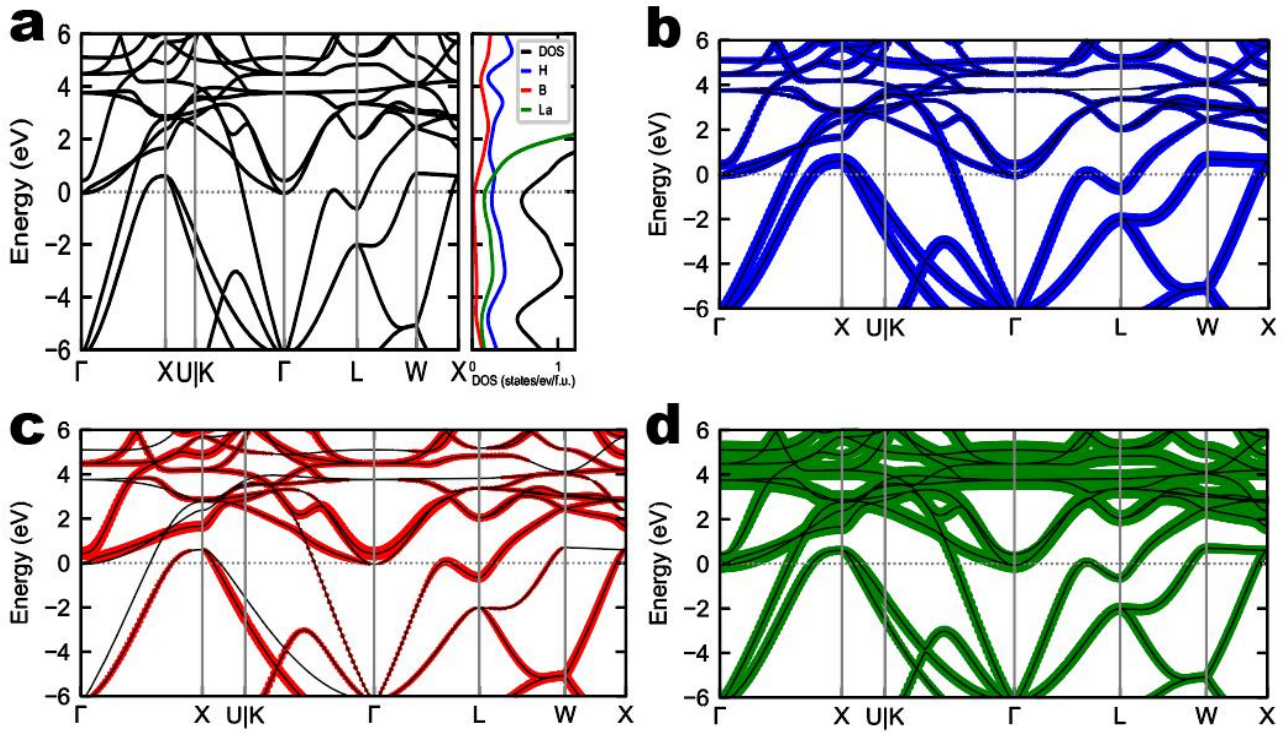

**FIG. S30.** Electronic structure of The of LaBH$_8$ at 100 GPa. **a** Electronic band structure (left panel) and projected density of states (right panel). And fatband of **b** H, **c** B and **d** Ca.

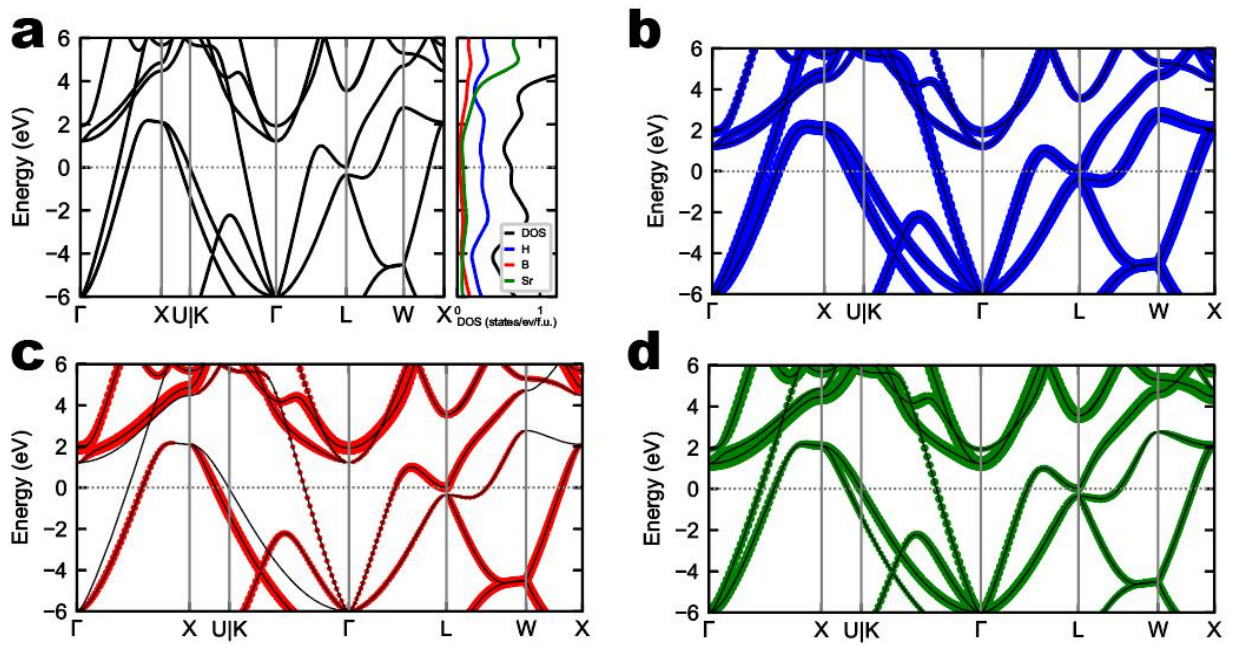

**FIG. S31.** Electronic structure of The of SrBH$_8$ at 200 GPa. **a** Electronic band structure (left panel) and projected density of states (right panel). And fatband of **b** H, **c** B and **d** Sr.

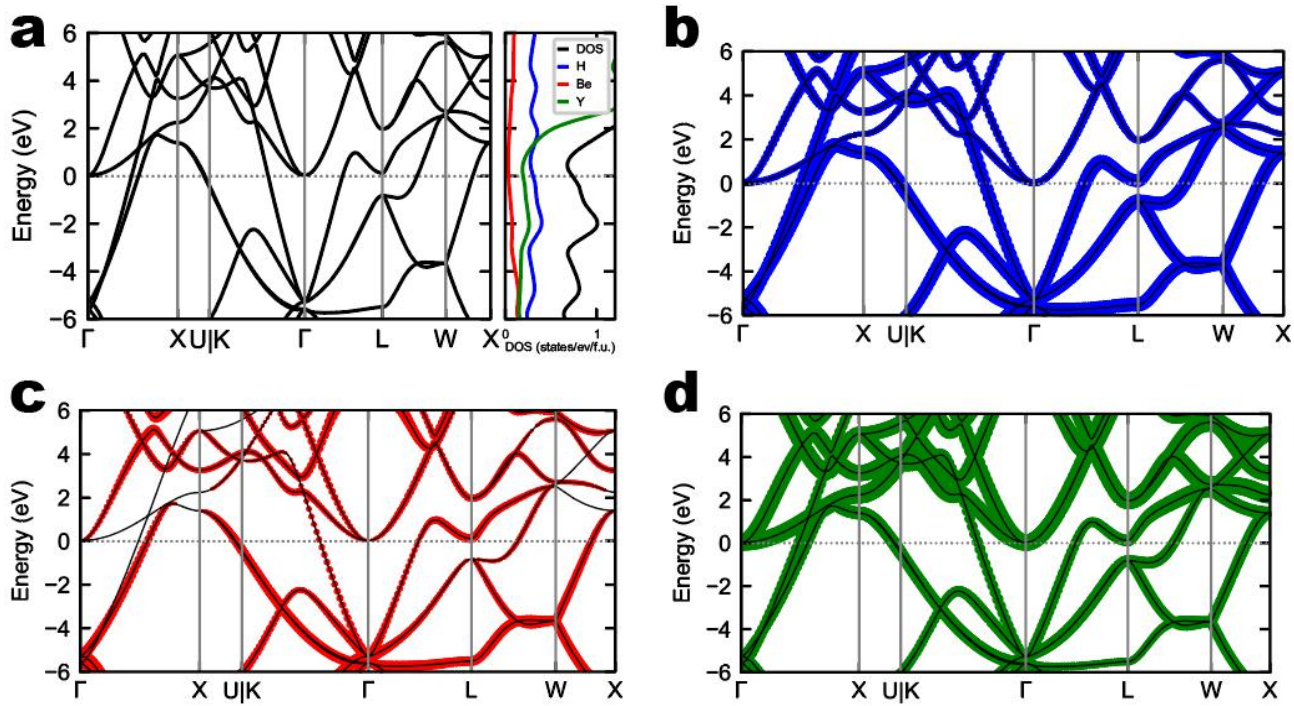

**FIG. S32.** Electronic structure of The of YBeH$_8$ at 100 GPa. **a** Electronic band structure (left panel) and projected density of states (right panel). And fatband of **b** H, **c** B and **d** Ca.

# 8. Phonon structure of "fluorite-like" hydrides

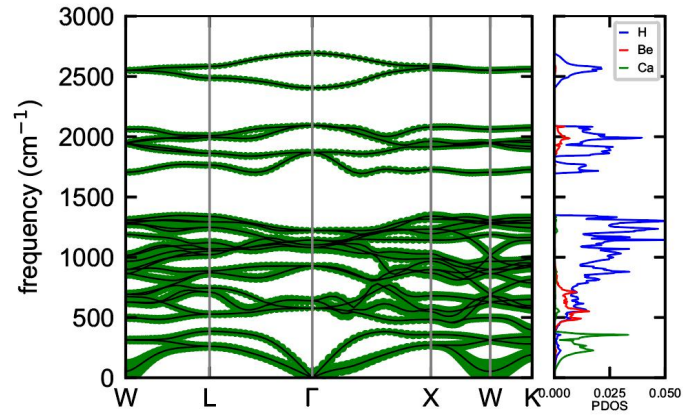

**FIG. S33.** Phonon structure of "fluorite-like" hydride CaBeH$_8$ at 210 GPa. Phonon structure with projective lambda (left panel) and phonon projected density of states (right panel).

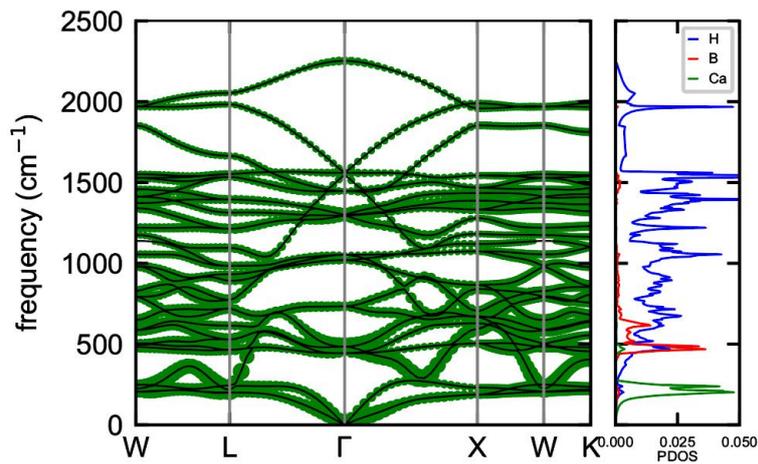

**FIG. S34.** Phonon structure of "fluorite-like" hydride CaBH$_8$ at 100 GPa. Phonon structure with projective lambda (left panel) and phonon projected density of states (right panel).

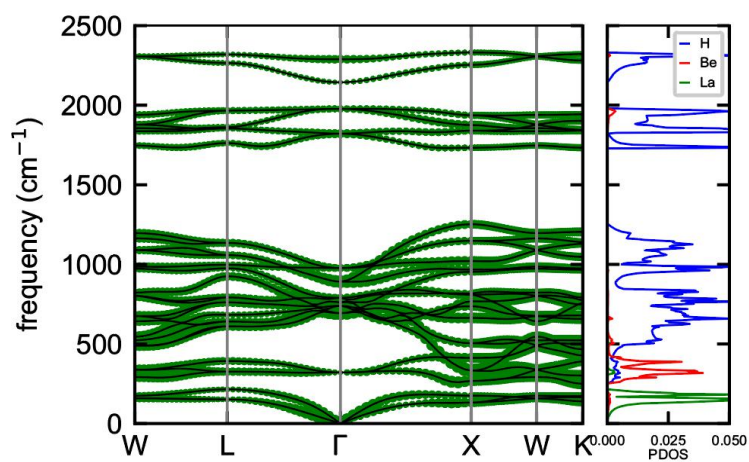

**FIG. S35.** Phonon structure of "fluorite-like" hydride LaAlH$_8$ at 100 GPa. Phonon structure with projective lambda (left panel) and phonon projected density of states (right panel).

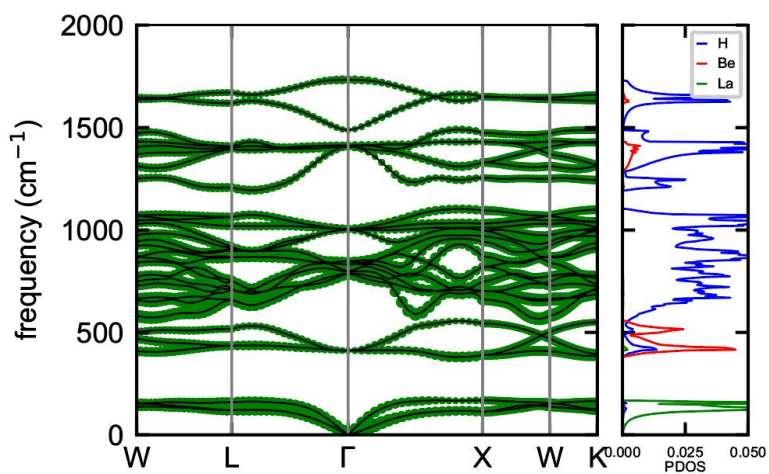

**FIG. S36.** Phonon structure of "fluorite-like" hydride LaBeH$_8$ at 50 GPa. Phonon structure with projective lambda (left panel) and phonon projected density of states (right panel).

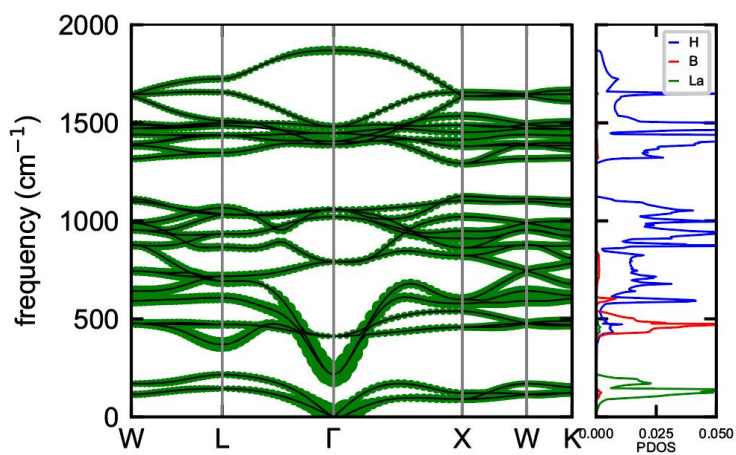

**FIG. S37.** Phonon structure of "fluorite-like" hydride LaBH$_8$ at 100 GPa. Phonon structure with projective lambda (left panel) and phonon projected density of states (right panel).

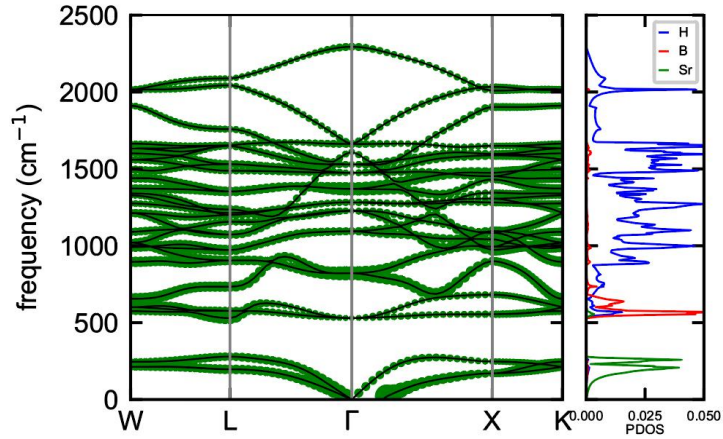

**FIG. S38.** Phonon structure of "fluorite-like" hydride SrBH$_8$ at 150 GPa. Phonon structure with projective lambda (left panel) and phonon projected density of states (right panel).

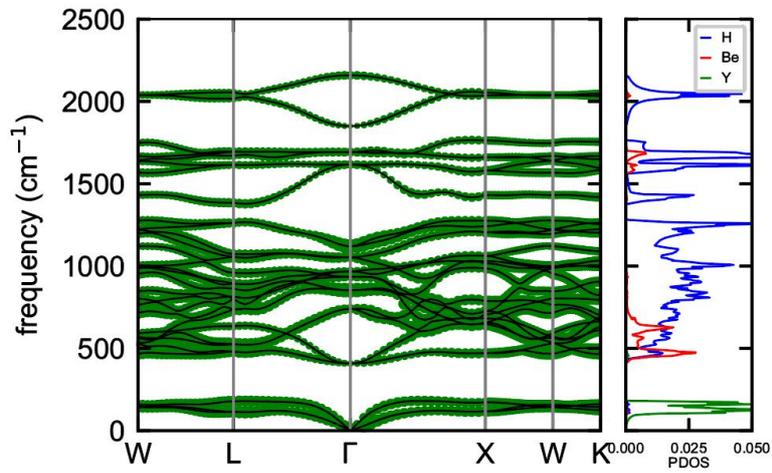

**FIG. S39.** Phonon structure of "fluorite-like" hydride YBeH$_8$ at 100 GPa. Phonon structure with projective lambda (left panel) and phonon projected density of states (right panel).

## 9. Superconductivity of "fluorite-like" backbone hydride

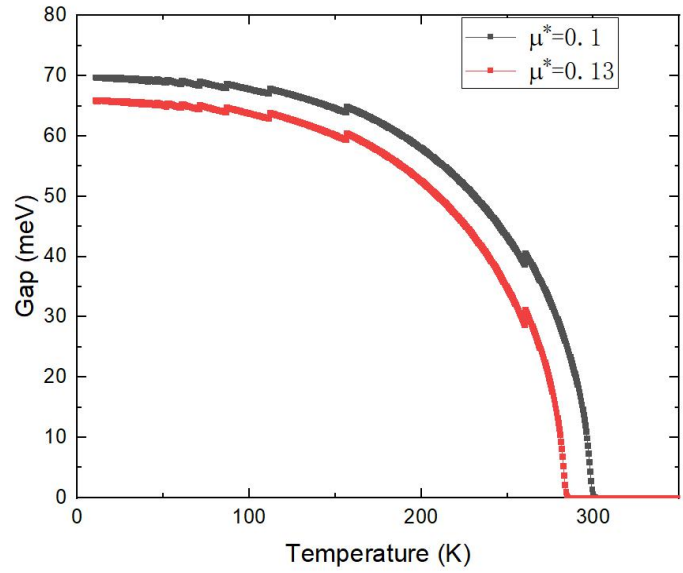

**FIG. S40.** Gap calculated by Eliashberg equation of "fluorite-like" hydride CaBeH$_8$ at 210 GPa.

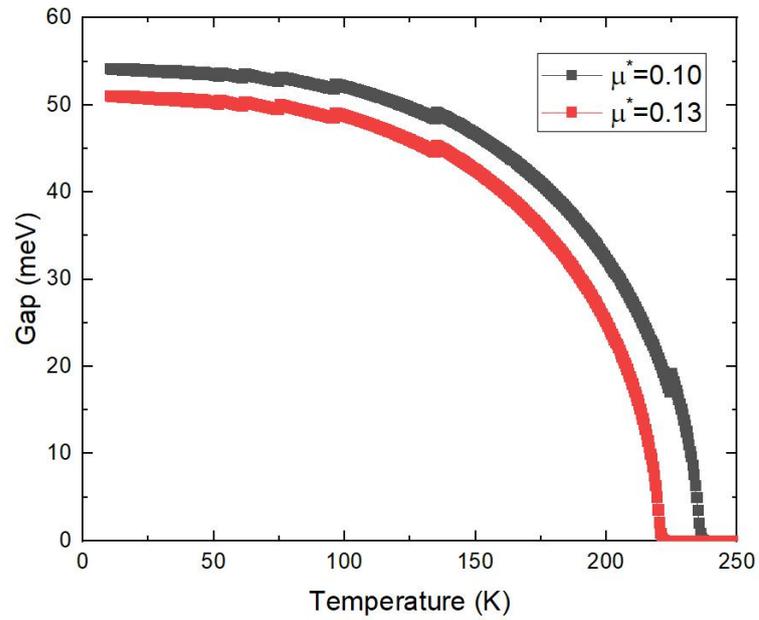

**FIG. S41.** Gap calculated by Eliashberg equation of "fluorite-like" hydride CaBH$_8$ at 100 GPa.

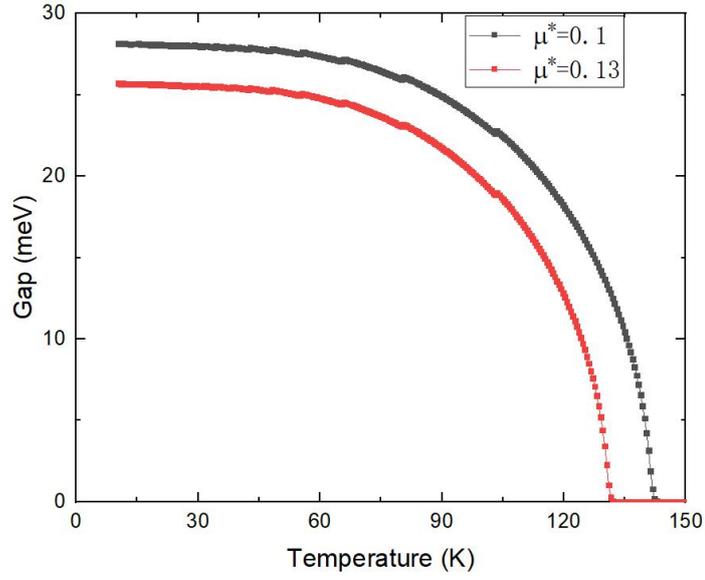

**FIG. S42.** Gap calculated by Eliashberg equation of "fluorite-like" hydride LaAlH$_8$ at 100 GPa.

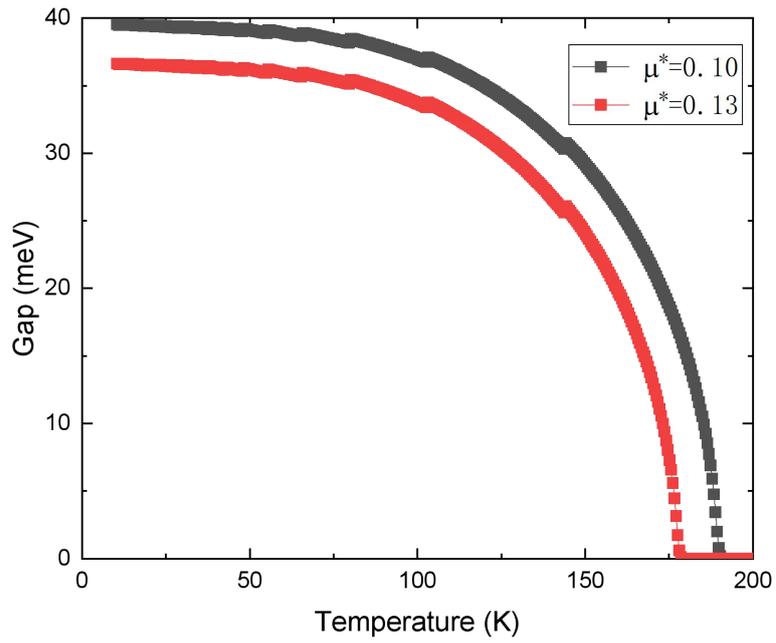

**FIG. S43.** Gap calculated by Eliashberg equation of "fluorite-like" hydride LaBeH$_8$ at 50 GPa.

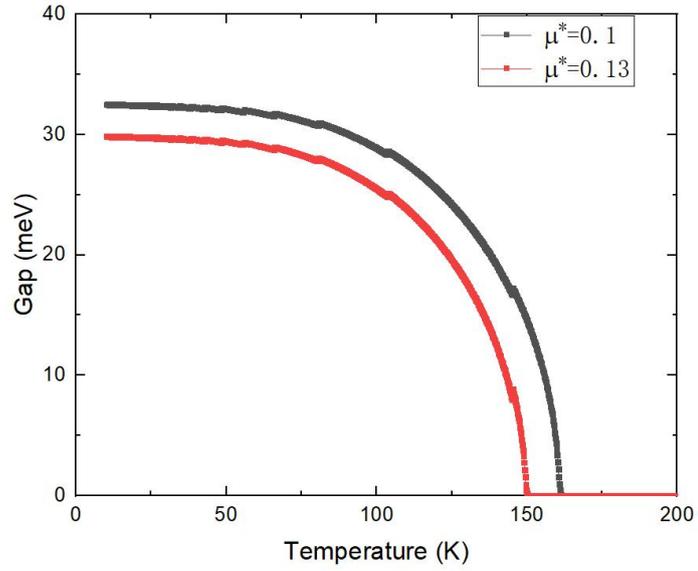

**FIG. S44.** Gap calculated by Eliashberg equation of "fluorite-like" hydride LaBH$_8$ at 100 GPa.

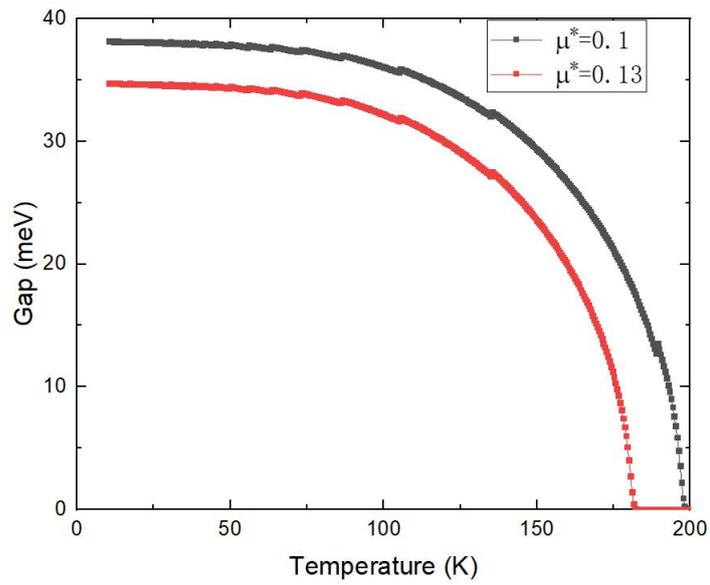

**FIG. S45.** Gap calculated by Eliashberg equation of "fluorite-like" hydride SrBH$_8$ at 150 GPa.

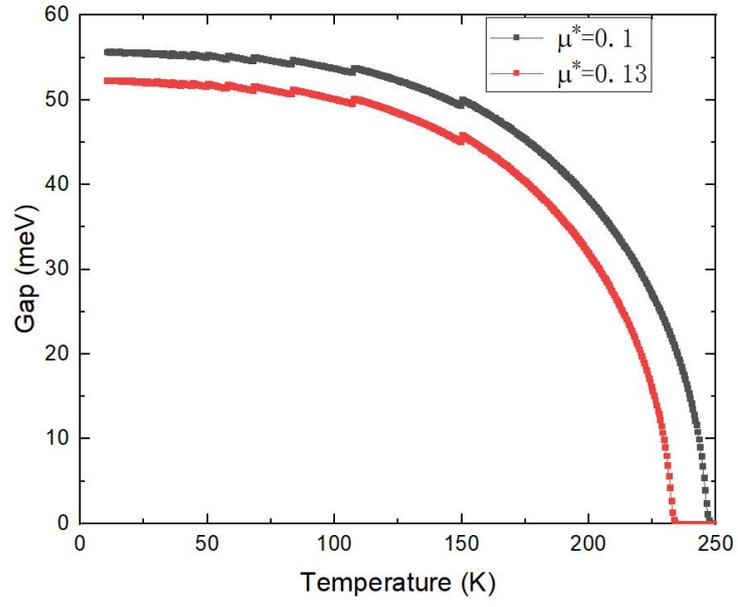

**FIG. S46.** Gap calculated by Eliashberg equation of "fluorite-like" hydride YBeH$_8$ at 100 GPa.

[1] C. J. Pickard and Needs, *Ab initio random structure searching,* J. Phys.-Condes. Matter, 053201 (2011).

[2] M. D. Segall, P. J. D. Lindan, M. J. Probert, C. J. Pickard, P. J. Hasnip, S. J. Clark, and M. C. Payne, *First-principles simulation: ideas, illustrations and the CASTEP code,* J. Phys.-Condes. Matter **14**, 2717, (2002).

[3] G. Kresse and J. Furthmüller, *Efficiency of ab-initio total energy calculations for metals and semiconductors using a plane-wave basis set,* Comput. Mater. Sci. **6**, 15, (1996).

[4] J. P. Perdew and Y. Wang, *Accurate and simple analytic representation of the electron-gas correlation energy,* Phys. Rev. B **45**, 13244, (1992).

[5] J. P. Perdew, K. Burke, and Y. Wang, *Generalized gradient approximation for the exchange-correlation hole of a many-electron system,* Phys. Rev. B **54**, 16533, (1996).

[6] G. Henkelman, A. Arnaldsson, and H. Jónsson, *A fast and robust algorithm for Bader decomposition of charge density,* Comput. Mater. Sci. **36**, 354, (2006).

[7] W. Tang, E. Sanville, and G. Henkelman, *A grid-based Bader analysis algorithm without lattice bias,* J. Phys.-Condes. Matter **21**, 084204, (2009).

[8] R. Dronskowski and P. E. Bloechl, *Crystal orbital Hamilton populations (COHP): energy-resolved visualization of chemical bonding in solids based on density-functional calculations,* J. Phys. Chem. **97**, 8617, (1993).

[9] V. L. Deringer, A. L. Tchougréeff, and R. Dronskowski, *Crystal Orbital Hamilton Population (COHP) Analysis As Projected from Plane-Wave Basis Sets,* J. Phys. Chem. A **115**, 5461, (2011).

[10] P. Giannozzi, S. Baroni, N. Bonini, M. Calandra, R. Car, C. Cavazzoni, D. Ceresoli, G. L. Chiarotti, M. Cococcioni, I. Dabo, A. Dal Corso, S. De Gironcoli, S. Fabris, G. Fratesi, R. Gebauer, U. Gerstmann, C. Gougoussis, A. Kokalj, M. Lazzeri, L. Martin-Samos, N. Marzari, F. Mauri, R. Mazzarello, S. Paolini, A. Pasquarello, L. Paulatto, C. Sbraccia, S. Scandolo, G. Sclauzero, A. P. Seitsonen, A. Smogunov, P. Umari, and R. M. Wentzcovitch, *QUANTUM ESPRESSO: A modular and open-source software project for quantum simulations of materials,* J. Phys.-Condes. Matter **21**, 395502, (2009).

[11] G. Kresse and D. Joubert, *From ultrasoft pseudopotentials to the projector augmented-wave method,* Phys. Rev. B **59**, 1758, (1999).

[12] C. R. and Dynes, *McMillan's equation and the Tc of superconductors,* Solid State Commun. **10**, 615, (1972).

[13] Hertel, *Transition temperature of strong-coupled superconductors,* Phys. Rev. B **167**, 331, (1968).

[14] G. M. Eliashberg, *Interactions between electrons and lattice vibrations in a superconductor,* Sov. Phys. **11**, 696, (1960).

[15] L. P. Gor'kov and V. Z. Kresin, *Pressure and high-Tc superconductivity in sulfur hydrides,* Sci. Rep. **6**, 25608, (2016).

[16] L. P. Gor'kov and V. Z. Kresin, *Colloquium: High pressure and road to room temperature superconductivity,* Rev. Mod. Phys. **90**, 16, (2018).

[17] P. B. Allen and R. C. Dynes, *Transition temperature of strong-coupled superconductors reanalyzed,* Phys. Rev. B **12**, 905, (1975).

[18] Z. Wang, Y. Yao, L. Zhu, H. Liu, T. Iitaka, H. Wang, and Y. Ma, *Metallization and superconductivity of $BeH_2$ under high pressure,* J. Chem. Phys. **140**, 124707, (2014).

[19] H. Wang, S. T. John, K. Tanaka, T. Iitaka, and Y. Ma, *Superconductive sodalite-like clathrate calcium hydride at high pressures,* Proc. Natl. Acad. Sci. U. S. A. **109**, 6463, (2012).

[20] Z. Shao, D. Duan, Y. Ma, H. Yu, H. Song, H. Xie, D. Li, F. Tian, B. Liu, and T. Cui, *Unique Phase Diagram and Superconductivity of Calcium Hydrides at High Pressures,* Inorg. Chem. **58**, 2558, (2019).

[21] C. H. Hu, A. R. Oganov, Q. Zhu, G. R. Qian, G. Frapper, A. O. Lyakhov, and H. Y. Zhou, *Pressure-induced stabilization and insulator-superconductor transition of BH,* Phys. Rev. Lett. **110**, 165504, (2013).

[22] I. Goncharenko, M. I. Eremets, M. Hanfland, J. S. Tse, M. Amboage, Y. Yao, and I. A. Trojan, *Pressure-induced hydrogen-dominant metallic state in aluminum hydride,* Phys. Rev. Lett. **100**, 045504, (2008).

[23] F. Peng, Y. Sun, C. J. Pickard, R. J. Needs, Q. Wu, and Y. M. Ma, *Hydrogen Clathrate Structures in Rare Earth Hydrides at High Pressures: Possible Route to Room-Temperature Superconductivity,* Phys. Rev. Lett. **119**, 107001, (2017).

[24] Y. Wang, H. Wang, J. S. Tse, T. Iitaka, and Y. Ma, *Structural morphologies of high-pressure polymorphs of strontium hydrides,* Phys. Chem. Chem. Phys. **17**, 19379, (2015).